\documentclass[]{interact}

\usepackage{epstopdf}% To incorporate .eps illustrations using PDFLaTeX, etc.
\usepackage[caption=false]{subfig}% Support for small, `sub' figures and tables

\usepackage{upgreek}

\usepackage[numbers,sort&compress,merge]{natbib}% Citation support using natbib.sty
\bibpunct[, ]{[}{]}{,}{n}{,}{,}% Citation support using natbib.sty
\renewcommand\bibfont{\fontsize{10}{12}\selectfont}% Bibliography support using natbib.sty

\theoremstyle{plain}% Theorem-like structures provided by amsthm.sty

\theoremstyle{definition}

\theoremstyle{remark}

\begin{document}

\articletype{FULL ARTICLE}% Specify the article type or omit as appropriate

\title{Channel-specific core-valence projectors for determining partial Auger
decay widths}

\author{
\name{Florian Matz\textsuperscript{a} and Thomas-Christian Jagau\textsuperscript{a}}
\affil{\textsuperscript{a}Division of Quantum Chemistry and Physical Chemistry, KU Leuven,
Celestijnenlaan 200F, 3001 Leuven, Belgium.}
}

\maketitle

\begin{abstract}
Auger decay is a relaxation process of core-vacant states in atoms and molecules, in 
which one valence electron fills the core vacancy while a second one is emitted. These 
states pose a challenge to electronic-structure theory, because they are embedded in 
the ionization continuum. Recently, we showed that molecular Auger decay can be described 
using complex-variable coupled-cluster (CC) methods and that partial widths and branching 
ratios can be computed based on a decomposition of the CC energy. Here, we introduce 
channel-specific core-valence projectors, dubbed Auger channel projectors, as a more 
general technique to evaluate partial widths from complex-variable treatments. We apply 
this new method to core-ionized states of neon, water, ammonia, and methane using CC 
singles and doubles (CCSD), equation-of-motion ionization potential CCSD, and 
configuration interaction singles (CIS) wave functions. Even though a single CIS 
calculation can never describe all Auger decay channels at once, we show that a 
combination of CIS calculations based on different reference states is able to 
recover partial and total decay widths from CC calculations to an excellent degree.

%by choosing appropriate reference states and combining their spin- and
%symmetry-incomplete expansions. Using complex-variable methods based on complex scaling of
%the basis functions, all methods agree on partial and total Auger decay widths of the
%core-ionized states of neon, water, ammonia and methane. 

%In this manuscript, we demonstrate that no high degree of correlation is necessary for the description of Auger decay: in fact, a description solely based on singly excited wave functions is possible using non-Hermitian quantum chemistry based on complex scaling of the Hamiltonian or the basis functions. 
%The results are satisfactory and the perspectives on applying the method to larger molecules and processes related to Auger decay such as intermolecular coulombic decay or electron-transfer-mediated decay are bright as the EOMIP-CCSD and CIS methods come at much lower computational costs than $\Delta$CCSD.
\end{abstract}


%%%%%%%%%%%%%%%%%%%%%%%%%%%%%%%%%%%%%%%%%%%%%%%%%%%%%%%%%%%%%%%%%%%%%%%%%%%%%%%%%%%%%%

\section{Introduction} \label{sec:int}

Auger decay is the most important relaxation process for core vacancies in systems 
with light nuclei~\cite{agarwal13}. It is common in systems irradiated by X-rays or
high-energy electrons where it causes secondary electron emission. These mechanisms 
can lead to complicated electronic and photonic spectra with many features that are 
of high interest for analytic methods~\cite{agarwal13}. However, spectra alone can only 
give limited information about the mechanisms at play when not supported by a theoretical 
description that allows the assignment of signals to different electronic transitions. 
Consequently, core-ionized states and Auger decay have attracted considerable interest 
in molecular electronic-structure theory. 

A fundamental challenge is that core-ionized states are electronic resonances embedded 
in the ionization continuum~\cite{jagau22}; their lifetime is finite precisely because 
of Auger decay. To model only the energy levels of the species involved in X-ray 
absorption or Auger spectra, that is, to reproduce the positions of the signals, it 
has been shown that the decay process can often be neglected. This can be done by means 
of the core-valence separation~\cite{cederbaum80} (CVS), which removes from the wave 
function those configurations that describe the decay. Electronic-structure methods 
used with CVS projectors include in particular algebraic diagrammatic construction 
(ADC)~\cite{wenzel14a,wenzel14b}, coupled-cluster (CC)~\cite{zheng19}, and 
equation-of-motion coupled-cluster (EOM-CC) methods~\cite{coriani15,vidal19,liu19}.

To actually describe Auger decay, one needs a more elaborate approach to treat the 
interaction of the metastable core-vacant state with the continuum. Corresponding 
approaches can be based on the R-matrix method~\cite{tennyson10,gorczyca00,garcia09} 
or the theory of electronic resonances by Fano~\cite{fano61} and Feshbach~\cite{feshbach62}. 
Here, the Hilbert space is partitioned into a bound and a continuum part and the 
outgoing Auger electron is represented either implicitly using Stieltjes 
imaging~\cite{langhoff74,carravetta87,averbukh05,kolorenc11,kolorenc20} or 
explicitly where its wave function is approximated as a plane wave, Coulomb wave 
or scattered molecular wave~\cite{ellison75,colle88,zaehringer92,yarzhemsky02,
inhester12,skomorowski21a,skomorowski21b}. These approaches have the advantage 
that electronic-structure methods for bound states can be used to describe 
the initial and final states of Auger decay, but modeling the wave function 
of the outgoing electron correctly can be challenging~\cite{zaehringer92,
skomorowski21b}.

Complex-variable techniques based on non-Hermitian quantum chemistry~\cite{
moiseyev11,jagau17,jagau22} offer an alternative framework for the treatment 
of Auger decay~\cite{matz22}. Here, electronic resonances are described as discrete 
$L^2$ integrable solutions of the Schr\"odinger equation with complex energy 
$E_\mathrm{res} = E - \mathrm{i} \Gamma/2$, where $E$ is the physical energy of 
the state and $\Gamma$ its decay width. In our earlier work~\cite{matz22}, we 
used complex scaling of the Hamiltonian~\cite{aguilar71,balslev71} and of the 
exponents of the basis functions~\cite{mccurdy78,moiseyev79} to describe molecular 
Auger decay. This allows a more black-box approach in the sense that no explicit 
treatment of the continuum is necessary but the computational cost is increased 
%&no assumptions for wave function of outgoing electrons?
because of unphysical parameter dependencies~\cite{jagau17,jagau22}. 

A special interest in the context of Auger decay lies in determining the probability 
with which a core-ionized state decays into a certain final state. This is usually 
quantified in terms of partial widths $\Gamma_i$, which add up to the total width 
$\Gamma$~\cite{agarwal13}. Partial widths determine the signal intensities in Auger 
spectra but they are also relevant because only certain final states of Auger 
decay can undergo further decay processes such as intermolecular Coulombic decay 
(ICD)~\cite{cederbaum97} or electron-transfer mediated decay (ETMD)~\cite{zobeley01}.

In our previous work~\cite{matz22}, we showed that one can decompose the imaginary 
part of the energy from a complex-variable CC singles and doubles (CCSD) calculation 
into contributions from different excitations that can be associated with the decay 
channels. This allowed the evaluation of partial decay widths from such calculations. 
However, a similar decomposition of the equation-of-motion ionization potential 
(EOMIP)-CCSD energy gave bad results. 

In the present work, we introduce Auger channel projectors (ACPs) as a more general 
technique to extract partial decay widths from complex-variable calculations. ACPs 
are similar to CVS projectors but applied only to individual decay channels. They 
enable accurate calculations of partial decay widths from CCSD, EOM-CCSD, and 
configuration interaction singles (CIS) wave functions and may potentially be 
applied to other wave functions as well. A more detailed discussion of the theory 
of ACPs is given in Sec. \ref{sec:th}, whereas Sec. \ref{sec:h2o} presents numerical 
results for core-ionized water and Sec. \ref{sec:appl} further applications to neon, 
ammonia, and methane. Some general conclusions are provided in Sec. \ref{sec:conc}.

\section{Theory} \label{sec:th}
\subsection{Describing Auger decay with complex-variable methods} \label{sec:thcv}

General overviews of complex scaling can be found, for example, in Refs.~\cite{
moiseyev11,jagau22}. Our approach to molecular Auger decay is based on the method 
of complex basis functions~\cite{mccurdy78,moiseyev79,white15}, which is a variant 
of complex scaling suitable for molecular electronic resonances. In this method, 
the exponents of several basis functions are multiplied by a complex number 
$\mathrm{e}^{-2\mathrm{i}\theta}$~with $\theta$ as scaling angle.  While it can 
be shown that the complex resonance energies $E_\mathrm{res} = E - \mathrm{i} 
\Gamma/2$ are independent of $\theta$ in the basis-set limit, $E_\mathrm{res}$ 
does depend on $\theta$ in actual calculations with a finite basis set. In 
accordance with previous work, we determine the optimal value of $\theta$ by 
minimizing $|\mathrm{d}E/\mathrm{d}\theta|$~\cite{braendas77,moiseyev78}. For 
this purpose, trajectories $E(\theta)$ need to be calculated. 

We discussed the application of complex-scaled methods to Auger decay in detail 
in our preceding work~\cite{matz22}. It is necessary to include in the wave 
function for a core-ionized state configurations in which the core vacancy 
is filled by one valence electron and another valence electron is emitted into 
the continuum. The latter is represented in complex-scaled methods by the 
excitation of an electron into a virtual orbital with a high real energy and 
a significant imaginary energy~\cite{matz22}. 

\begin{figure}[ht]
\centering
\includegraphics[width=\linewidth]{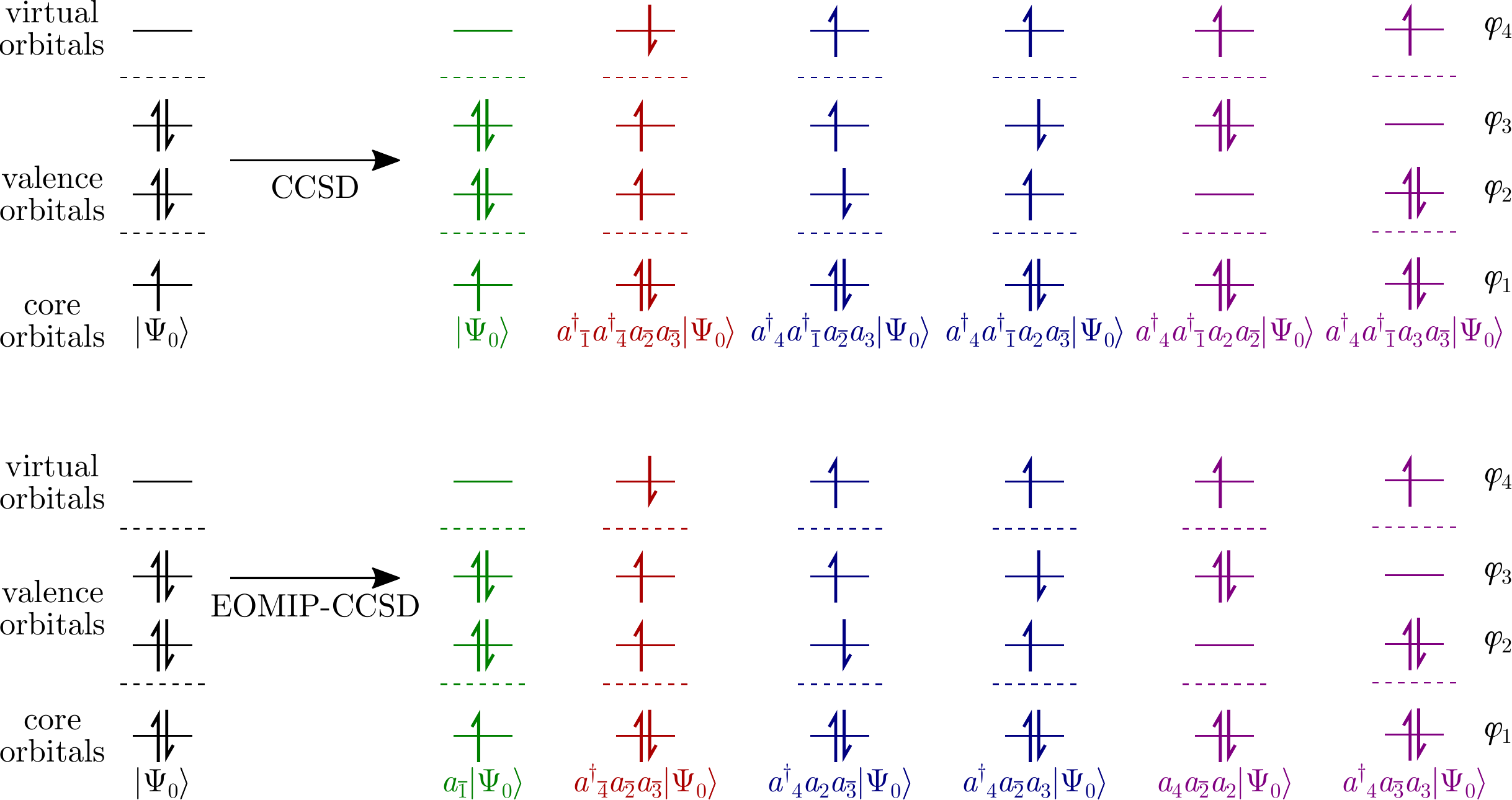}
\caption{Determinants needed to describe Auger decay in a Hilbert space spanned by 
4 orbitals as obtained from CCSD (upper part) or EOMIP-CCSD (lower part) calculations. 
See text for further explanation.}
\label{fig:ccsd} \end{figure}

We showed~\cite{matz22} that $\Delta$CCSD~\cite{cizek66,cizek69,shavitt09} and 
EOMIP-CCSD~\cite{nooijen93,stanton93,shavitt09} yield accurate total and partial 
Auger decay widths and discussed the strengths and weaknesses of these methods. 
In the $\Delta$CCSD approach, the neutral ground state of the system and the 
core-ionized decaying state are described independently from each other by two 
CCSD wave functions. All final states of Auger decay are double excitations with 
respect to the core-ionized reference determinant $|\Psi_0\rangle$ and thus included 
in the CCSD wave function. This is illustrated for an exemplary four-orbital system 
in the upper panel of Figure~\ref{fig:ccsd}. While all determinants included in 
this wave function have $M_S =  1/2$, a further distinction is possible when not 
taking into account the outgoing electron in the virtual orbital: There are 
closed-shell determinants (marked in purple), high-spin determinants (marked 
in red) that represent a triplet final state, and low-spin open-shell determinants 
(marked in blue) that can be linearly combined to form either a singlet or a 
triplet final state. Only when neglecting the outgoing electron is the description 
of the final states spin-complete. 

An alternative approach is EOMIP-CCSD, where the wave function of the core-ionized
state is constructed by applying a linear excitation operator to a CCSD wave 
function for the neutral ground state. In EOMIP-CCSD, this operator consists 
of 1-hole and 2-hole-1-particle excitations. As illustrated in the lower panel 
of Figure~\ref{fig:ccsd}, all Auger decay channels can be written as 2-hole-1-particle 
excitations starting from the neutral ground state and are thus included in
the EOMIP-CCSD wave function. The same classification of the decay channels 
as in CCSD is possible and represented in Figure~\ref{fig:ccsd} by the same 
colors. However, in contrast to an open-shell CCSD approach, the EOMIP-CCSD 
wave function is by construction a spin eigenfunction. 

In our previous work~\cite{matz22}, we demonstrated that a main advantage of 
$\Delta$CCSD over EOMIP-CCSD is that partial decay widths can be evaluated from 
contributions $\mathrm{Im}\big[ t_{ij}^{ab} \, \langle ij||ab\rangle \big]$ to 
the imaginary part of the energy where $i$ and $j$ are valence orbitals, $a$ is 
the core hole, and $b$ is any virtual orbital representing the state of the 
outgoing Auger electron. For an EOMIP-CCSD wave function, a similar energy 
decomposition leads to artificially large partial widths even though total 
decay widths are not less accurate than those calculated with CCSD~\cite{matz22}. 

%%%%%%%%%%%%%%%%%%%%%%%%%%%%%%%%%%%%%%%%%%%%%%%%%%%%%%%%%%%%%%%%%%%%%%%%%%%%%%%%%%%%%%

%Core-ionized states are thus often described with expensive correlation methods 
% such as (equation of motion-)coupled-cluster with singles and
% doubles~\cite{cizek66,cizek69,shavitt09,emrich81,sekino84,stanton93,skomorowski21a,
% skomorowski21b,matz22,zheng19,park19,frati19,ghosh17}, Algebraic Diagrammatic Construction
% (ADC)~\cite{schirmer82,tarantelli94,averbukh05,kolorenc11,kolorenc20,wenzel14a,wenzel14b}
% or Configuration Interaction (CI)~\cite{inhester12,inhester14,peng16,zhang12}. 

% The high computational cost of these state-of-the-art methods restricts the size of 
% molecules that can be treated.

% Decisive for the description of Auger decay is the choice of the wave function
%parametrization. The well-known challenge of electron correlation needs to be solved in
%order to achieve an accurate representation of the electronic density and energy, which is
%complicated even more by the strong perturbation of the potential that core-ionization
%elicits and the highly excited character of core-ionized states.

%%%%%%%%%%%%%%%%%%%%%%%%%%%%%%%%%%%%%%%%%%%%%%%%%%%%%%%%%%%%%%%%%%%%%%%%%%%%%%%%%%%%%%

\subsection{Auger channel projection}\label{sec:thacp}

In this work, we demonstrate how to gain access to partial decay widths using other 
complex-variable electronic-structure methods besides CCSD. The new method can be 
seen as a generalization of the CVS projector~\cite{cederbaum80}. The latter removes 
from the excitation manifold all determinants where valence electrons are excited 
into core orbitals, thus all those marked in red, blue and purple in Figure~\ref{fig:ccsd}. 
In this way, the decay is quenched and the imaginary part of the energy stays zero 
apart from contributions resulting from the representation of the wave function in 
a finite basis set.\cite{jagau17,jagau22}

The CVS projector can be thought of as a sum of projectors representing individual 
Auger decay channels, which we will call Auger channel projectors (ACPs). We now 
determine the wave function under the constraint that only the amplitudes leading 
to certain Auger final states are zero. Because we would like to keep the wave function 
as similar as possible to those obtained with or without CVS, we suggest two ways to 
use this concept to determine partial widths:

In the ``in'' variant of the ACP, a single Auger decay channel is included in the 
wave function so that the imaginary part of the energy changes from approximately 
zero in the CVS method to a negative number. Because the imaginary part of the 
energy is not necessarily zero in finite basis set calculations with a CVS method, 
we choose to calculate the partial width as the difference between the imaginary 
energies obtained with ACP and CVS. The advantage of this ``in'' variant is that 
the observed decay can be attributed unequivocally to the active decay channel.

In the ``out'' variant of the ACP, a single decay channel is removed from the 
excitation manifold. Im($E$) will be less negative than with the original method 
where all decay channels are active. We then evaluate the partial width from the 
energy difference between the unmodified wave function and the ACP variant with 
one channel projected out. This carries the benefit that the result is much closer 
to the wave function that includes all decay channels so that the interaction between 
different decay processes and its effect on the energy is considered.

Since the computational cost of a single EOMIP-CCSD calculation scales as $N_\mathrm{occ}^2 
N_\mathrm{virt}^3$ and since there are in general $N_\mathrm{occ}^2$ decay channels, 
an ACP-EOMIP-CCSD treatment effectively scales as $N_\mathrm{occ}^4 N_\mathrm{virt}^3$ 
if all decay channels shall be considered individually. The cost of CCSD scales as 
$N_\mathrm{occ}^2 N_\mathrm{virt}^4$ so that ACP-EOMIP-CCSD is, in theory, preferable 
over decomposition of the CCSD energy in terms of computational cost if $N_\mathrm{virt} 
> N_\mathrm{occ}^2$. This is the case in typical complex-variable calculations. Also, 
EOMIP-CCSD can be based on a closed-shell reference state, which requires less memory 
than an open-shell CCSD treatment of the core-ionized state. We note, however, that our 
current implementation of ACPs is not able to carry out multiple consecutive EOMIP-CCSD 
calculations with different ACPs; instead the CCSD calculation for the reference state 
needs to be repeated. 

%This procedure is not limited to certain wave function parametrizations -- for CCSD, the results are similar to those that equation~\ref{eq:ccdecomp} produces, but it was shown that the results for partial Auger decay widths in EOMIP-CCSD are unreliable. The reasons for this failure are also arising from the product of multiple excitation operators, as for example $\hat{L}^\dagger\cdot\hat{R}$ in the EOMIP-CCSD wave function. 

%%%%%%%%%%%%%%%%%%%%%%%%%%%%%%%%%%%%%%%%%%%%%%%%%%%%%%%%%%%%%%%%%%%%%%%%%%%%%%%%%%%%%%

\subsection{Application of Auger channel projectors to CIS wave functions} \label{sec:cistheory}

Because it involves two electrons, Auger decay is sometimes referred to as a
``correlation-driven'' process~\cite{bauch12,pisanty14,driver20}. However, this 
does not mean that a correlated electronic-structure method needs to be employed 
for describing it. Namely, it is possible to describe Auger decay with a CIS wave 
function~\cite{delbene71,dreuw05}, which only comprises single excitations out of 
some reference determinant $|\Psi_0\rangle$. CIS can be implemented with $N^4$ 
scaling and is size-extensive similar to EOM-CC methods and in contrast to 
higher-order CI methods.  

\begin{figure}
\centering
\includegraphics[width=.9\linewidth]{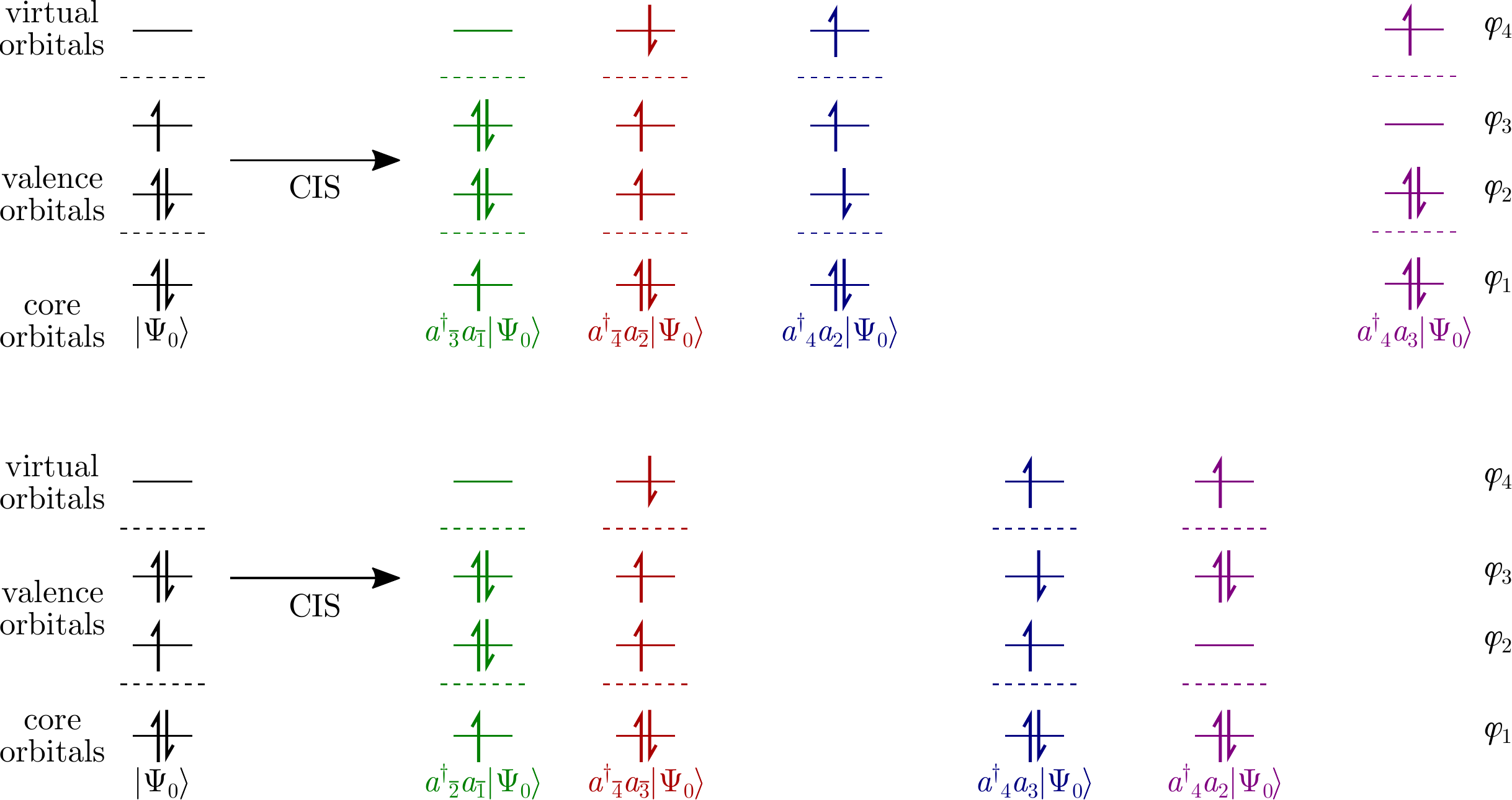}
\caption{Determinants needed to describe Auger decay in a Hilbert space spanned 
by 4 orbitals as obtained from different CIS calculations. See text for further 
explanation.}
\label{fig:cis}
\end{figure}

Describing Auger decay with a CIS wave function requires to choose a reference 
determinant whose single-excitation manifold comprises both the core-ionized 
determinant and the final state of Auger decay. As illustrated in Figure \ref{fig:cis}, 
this is only possible if the reference determinant has a hole in a valence shell 
and this hole is also present in the Auger final state. This implies that a CIS 
calculation can only describe a subset and never all possible Auger decay channels. 
The imaginary part of the CIS energy must therefore not be interpreted as a total 
Auger decay width. 

Figure~\ref{fig:cis} shows that each decay channel is described by at least one 
CIS wave function. More specifically, determinants representing closed-shell 
final states (purple in Figure~\ref{fig:cis}) are included in only one CIS wave 
function, whereas the red determinants associated with triplet final states are 
are each contained in exactly two CIS wave functions. Open-shell singlet final 
states are described by the blue determinants, which are also each contained in 
exactly one CIS wave function. However, for a proper description of an open-shell 
singlet, one needs to combine the blue determinants in pairs while this is not 
needed for the red ones. Also note that the blue determinants do not contribute 
to the low-spin components of the triplet final states because all determinants 
in Figure \ref{fig:cis} have $M_S=1/2$ and the outgoing Auger electron has 
necessarily alpha spin in the blue determinants. 

Even though this analysis is approximate because we start from an unrestricted 
Hartree-Fock (HF) reference determinant, which is not a spin eigenfunction, it 
suggests the following procedure to compute the total Auger decay width using 
CIS: We first construct all possible HF determinants with a hole in a valence 
orbital and then apply the ACPs as introduced in Sec. \ref{sec:thacp} to the 
corresponding CIS wave functions to identify the contributions of individual 
determinants to the decay width. In a second step, we combine the obtained 
results as follows: Contributions from triplet final states (red determinants 
in Figure~\ref{fig:cis}) are averaged, while contributions from singlet final 
states (blue and purple determinants in Figure~\ref{fig:cis}) are simply added up. 
We add that a straightforward summation of the imaginary energies obtained in the 
different CIS calculations may offer a good approximation to the total Auger decay 
width because triplet decay channels typically contribute much less than singlet 
states. 

We also note that there is no unambiguous way to obtain the core-ionization energy 
from such CIS calculations because the CIS eigenvalues represent only energy differences 
between the core-ionized state and different valence-ionized states and do not involve 
the neutral ground state. Furthermore, these energies cannot be expected to be accurate 
given the limited accuracy of CIS excitation energies in general~\cite{dreuw05}. 

\subsection{Special considerations for non-Abelian point groups}\label{sec:symtheory}

Molecules with non-Abelian point-group symmetry can possess degenerate occupied 
orbitals, which complicates the extraction of partial decay widths from complex-variable 
calculations. We illustrate this in the following for methane (CH$_4$) which has 
the electronic configuration (1a$_1$)$^2$(2a$_1$)$^2$(1t$_2$)$^6$. The analysis 
applies to CIS, CCSD, and EOMIP-CCSD wave functions alike. 

As discussed in Sec.~\ref{sec:thacp}, it is necessary to set to zero all excitations 
that describe a particular decay channel in order to calculate the respective partial 
width. In decay channels where at most one degenerate occupied orbital is involved, 
these excitations can be identified by considering only the occupancy of the valence 
orbitals. 

For example, for methane every (2a$_1$)$^{-2}$ determinant describes decay into the 
corresponding $^1$A$_1$ state. The outgoing Auger electron must occupy a virtual orbital 
of a$_1$ symmetry as well because the core-ionized state has $^2$A$_1$ symmetry overall 
but this need not be taken into account for the determination of partial widths. Likewise, 
every (2a$_1$)$^{-1}$(1t$_2$)$^{-1}$ determinant contributes to decay into the $^1$T$_2$ 
or $^3$T$_2$ state depending on the spin of the outgoing electron. Here, the equivalence 
of (2a$_1$)$^{-1}$(2t$_2^x$)$^{-1}$, (2a$_1$)$^{-1}$(2t$_2^y$)$^{-1}$, and 
(2a$_1$)$^{-1}$(2t$_2^z$)$^{-1}$ can be exploited to reduce the number of necessary
calculations. 

In contrast, (1t$_2$)$^{-2}$ determinants contribute to four different decay channels: 
$^1$A$_1$, $^1$T$_2$, $^1$E, and $^3$T$_1$ according to the direct product t$_2$ 
$\otimes$ t$_2$. Singlet and triplet channels can be distinguished depending on 
whether the outgoing electron has $\alpha$ or $\beta$ spin, but for the distinction 
between the different singlet states it is necessary to consider in addition if the 
irreducible representation of the virtual orbital where the outgoing electron 
resides is a$_1$, e, or t$_2$. The partial widths can then be determined by 
projecting out at once all (1t$_2$)$^{-2}$ determinants where virtual orbitals 
of a particular symmetry have been filled and bearing in mind that the irreducible 
representation of the core-ionized state is A$_1$.

We modified our previous implementation~\cite{matz22} according to these thoughts; 
however, special attention must be paid to the case of CIS. As discussed in Sec. 
\ref{sec:cistheory}, CIS calculations need to be based on HF determinants with a 
hole in a valence shell. If this shell is degenerate in the closed-shell HF wave 
function, the degeneracy is broken in the corresponding cationic wave function. 
As a consequence, the degeneracy of virtual shells is also lost so that the 
identification of the physical decay channels becomes ambiguous.

In this work, we assign the symmetry-broken virtual orbitals that host the Auger 
electron to the irreducible representations of the molecular point group by 
energetic proximity: those which are closest to each other are assumed to be 
degenerate if the electronic wave function transformed to the molecular point 
group. Consider as a numerical illustration a HF wave function for the methane 
cation: The artificial splitting in the virtual e orbitals induced by an 
asymmetric occupation of the $1\mathrm{t}_2$ orbitals amounts to at most 0.07 a.u., 
which is considerably lower than the energetic separation between virtual orbitals 
belonging to different shells. The identification of the decay channels is thus easy. 
However, this is not always the case, especially not if the virtual orbitals to be 
distinguished belong to the same irreducible representation as the valence shell 
with the hole. As an example, in the ammonia cation (NH$_3^+$) the artificial 
splittings in the virtual e orbitals amount to up to 0.5 a.u. which makes a 
distinction between the $^1$E and $^1$A$_1$ decay channels dubious.

\section{Auger decay of core-ionized water} \label{sec:h2o}

As a proof of concept, we use the 1a$_1^{-1}$ state of the water molecule, with an
electronic configuration of (1a$_1$)$^1$(2a$_1$)$^2$(1b$_2$)$^2$(3a$_1$)$^2$(1b$_1$)$^2$.
We showed before that Auger decay in this system can be accurately described using 
a basis where s and p shells are taken from the aug-cc-pCV5Z basis set and the shells 
with higher angular momentum from the aug-cc-pCVTZ basis set. Two complex-scaled s-, 
p-, and d-shells were added on top at each atom~\cite{matz22}. As the augmentation 
of the basis set with diffuse shells was found to be of minor importance, we employ 
the ``cc-pCVTZ (5sp) + 2$\times$(spd)'' basis set in this work. Additional results 
using 4 complex-scaled s, p and d shells are reported in the Supplementary Material. 
The exponents of the complex-scaled shells are obtained using the strategy published earlier~\cite{matz22} and available from the Supplementary Material. We conducted all
calculations using a modified version of the Q-Chem program package~\cite{epifanovsky21}.
The CIS calculations were performed with the complex-variable CCSD and EOM-CCSD codes of 
Q-Chem~\cite{bravaya13,zuev14,white15,white15b,white17} where we set to zero all 
CCSD amplitudes and the EOM-CCSD doubles amplitudes. 

\subsection{CCSD and EOMIP-CCSD results}\label{sec:acpresults}

The total decay half-width of water, calculated as the difference between the 
imaginary energies of the ground state and the core-ionized state, amounts to 
71 meV with EOMIP-CCSD at $\theta_\mathrm{opt} = 22^\circ$ and 70 meV with 
$\Delta$CCSD at $\theta_\mathrm{opt} = 16^\circ$. Partial half-widths obtained 
using the energy decomposition method introduced in our earlier work~\cite{matz22} 
are shown in columns 3 and 4 of Table~\ref{tab:h2opw}. These results illustrate 
that energy decomposition works only for CCSD but not for EOM-CCSD; the reasons 
were discussed in Ref.~\cite{matz22}.

\begin{table}
\tbl{Partial decay half-widths of core-ionized water calculated with CCSD 
and EOMIP-CCSD using the cc-pCVTZ (5sp) basis set with 2 complex scaled 
s-, p- and d-shells on each atom. All values in meV.}
{\begin{tabular}{lrrrrrrr}\toprule
Decay & EOM-CCSD & CCSD & EOMIP-CCSD & \multicolumn{2}{c}{ACP-CCSD} & 
\multicolumn{2}{c}{ACP-EOMIP-CCSD} \\
channel & Fano\textsuperscript{a} & \multicolumn{2}{c}{decomposition} & 
``out'' & ``in'' & ``out'' & ``in'' \\ \midrule
2a$_1$2a$_1$ & 7.7 & 9.1 & 8.1 & 9.8 & 8.9 & 8.5 & 8.4 \\
2a$_1$3a$_1$ (singlet) & 6.8 & 6.8 & 2.3 & 6.6 & 6.7 & 6.6 & 6.3 \\
2a$_1$3a$_1$ (triplet) & 1.9 & 1.3 & 3.4 & 1.3 & 0.9 & 1.2 & 1.0 \\
2a$_1$1b$_1$ (singlet) & 4.8 & 6.5 & --3.7 & 6.0 & 5.4 & 6.1 & 5.9 \\
2a$_1$1b$_1$ (triplet) & 2.1 & 1.6 & 2.4 & 1.5 & 1.1 & 1.4 & 1.2 \\
2a$_1$1b$_2$ (singlet) & 3.2 & 4.1 & 3.5 & 3.8 & 3.2 & 4.2 & 3.9 \\
2a$_1$1b$_2$ (triplet) & 1.5 & 1.0 & 3.1 & 1.1 & 0.7 & 1.0 & 0.8 \\
3a$_1$3a$_1$ & 4.5 & 5.9 & 15.1 & 5.1 & 5.0 & 4.6 & 4.7 \\
3a$_1$1b$_1$ (singlet) & 6.4 & 8.3 & 23.2 & 7.3 & 7.1 & 6.9 & 6.8 \\
3a$_1$1b$_1$ (triplet) & 0.3 & 0.1 & 0.2 & 0.1 & 0.1 & 0.1 & 0.1 \\
3a$_1$1b$_2$ (singlet) & 4.8 & 8.3 & 22.4 & 7.2 & 6.9 & 6.1 & 6.0 \\
3a$_1$1b$_2$ (triplet) & 0.2 & 0.1 & 0.2 & 0.1 & 0.1 & 0.1 & 0.1 \\
1b$_1$1b$_1$ & 6.7 & 8.1 & 20.7 & 7.1 & 7.1 & 6.4 & 6.5 \\
1b$_1$1b$_2$ (singlet) & 5.4 & 6.9 & 21.5 & 6.0 & 5.9 & 5.8 & 5.8 \\
1b$_1$1b$_2$ (triplet) & 0.0 & 0.0 & 0.0 & 0.0 & 0.0 & 0.0 & 0.0 \\
1b$_2$1b$_2$ & 3.6 & 5.0 & 14.2 & 4.3 & 4.1 & 3.9 & 3.8 \\\midrule
Sum & 59.9 & 73.1 & 136.7 & 67.3 & 62.4 & 62.8 & 61.1 \\\bottomrule
\end{tabular}}
\tabnote{\textsuperscript{a} From reference~\cite{skomorowski21b}.}
\label{tab:h2opw}
\end{table}

In contrast, the Auger channel projectors introduced in Sec. \ref{sec:thacp} 
work for both, CCSD and EOMIP-CCSD. Results obtained with the ``in'' and 
``out'' versions are shown in the last four columns of Table \ref{tab:h2opw}, 
they are very similar to those obtained by decomposition of the CCSD energy 
and in qualitative agreement with values obtained using the Feshbach-Fano 
method~\cite{skomorowski21a,skomorowski21b}. The root mean square (RMS) 
deviations with respect to the former results vary from 0.4 meV (``out''-ACP-CCSD) 
to 1.0 (``in''-ACP-EOMIP-CCSD) meV. We add that all ACP calculations were 
performed at the $\theta_\mathrm{opt}$ of the respective unmodified method; 
we did not explore the impact of optimizing $\theta$ for each channel. 

The results from Table \ref{tab:h2opw} validate the ACP approach. The ``in'' 
and ``out'' approaches give almost identical values, the RMS deviations between 
them are only 0.4~meV~(CCSD) and 0.16~meV (EOMIP-CCSD), respectively, where the 
``out'' method typically yields partial widths closer to the energy decomposition 
approach. The largest deviations from energy decomposition occur in the high-energy 
singlet channels (1b$_1$1b$_1$, 3a$_1$1b$_1$, and 3a$_1$1b$_1$) where ACPs yield 
partial half-widths that are lower by about 1 meV. Overall, the differences between
ACPs and energy decomposition are even smaller. In fact, the comparison with partial 
widths of H$_2$O reported in our previous work~\cite{matz22} shows that adding diffuse
functions or changing details of the energy decomposition procedure make a larger 
impact.

\subsection{CIS results}\label{sec:cisresults}

As discussed in Sec.~\ref{sec:cistheory}, one needs to perform multiple separate 
CIS calculations to consider all decay channels. In the case of H$_2$O, they are 
based on four different cationic HF references (2a$_1^{-1}$, 1b$_2^{-1}$, 3a$_1^{-1}$, 
1b$_1^{-1}$) where we used the maximum overlap method~\cite{gilbert08} to ensure 
convergence to the right state. 

To determine the optimal value of the scaling angle $\theta$, we considered two 
different strategies. A straightforward procedure is to determine separate values 
for $\theta$ for each of the four CIS wave functions by minimizing 
$|\mathrm{d}E_\mathrm{CIS} / \mathrm{d}\theta|$. In the present case, this results 
in values between 14 and 28$^\circ$. However, our ACP-CCSD and ACP-EOMIP-CCSD 
results (see Sec. \ref{sec:acpresults}) suggest that it is not necessary to use 
a different $\theta$ for each wave function. The partial widths of all decay 
channels are determined at the same $\theta$ in these calculations and still 
sum up to an accurate total width. This is why we additionally used the sum 
$\sum_{i=1}^4 |\mathrm{d}E_\mathrm{CIS}^i / \mathrm{d}\theta|$ as criterion 
to find a uniform $\theta$ for all calculations. This yields $\theta_\mathrm{opt} = 
28^\circ$; at this value each trajectory exhibits a local minimum in 
$|\mathrm{d}E_\mathrm{CIS} / \mathrm{d}\theta|$, which is, however, not always 
the global minimum. One may speculate that the occurrence of multiple minima 
in $|\mathrm{d}E_\mathrm{CIS}/\mathrm{d}\theta|$ is an artifact stemming from 
the unphysical restriction of the configuration space. 

Our calculations show that the numerical differences between these two procedures 
are very small. In the following, we concentrate on the results obtained using a 
uniform $\theta$, whereas results obtained with a separate $\theta$ for each CIS 
calculation are available in the Supplementary Material. 

The last row of Table~\ref{tab:h2opwcis} shows that the imaginary parts of the CIS 
energies of the core-ionized states vary between 16 meV and 25 meV and are thus 
much lower than the CCSD value (70 meV). This is a consequence of the fact that 
each of the four CIS wave functions contains only a subset of the decay channels 
present in the CCSD wave function. The simplest way to solve this problem is to 
add up the Im($E$) values from the four CIS calculations, which results in a 
total half-width of 77.5~meV, that is, the deviation from CCSD is ca. 10\%. As 
discussed in Sec.~\ref{sec:cistheory}, this simple summation counts the triplet 
channels twice. This is confirmed by adding up the CCSD partial widths (last 
column of Table~\ref{tab:h2opwcis}) and doubling the widths of the triplet 
channels, which results in 77.2 meV. 

\begin{table}
\tbl{Partial decay half-widths of core-ionized water calculated with CIS using 
different reference determinants and the ``out'' version of ACPs. The cc-pCVTZ (5sp) 
basis set with 2 complex scaled s-, p- and d-shells on each atom and a uniform 
$\theta = 28^\circ$ are used. Results from the decomposition of the CCSD energy 
(taken from Table \ref{tab:h2opw}) are given as well. All values in meV.}
{\hspace{1.5cm}\begin{tabular}{lrrrrrr}\toprule
Decay & \multicolumn{5}{c}{CIS} & CCSD \\
channel & 2a$_1^{-1}$ & 1b$_2^{-1}$ & 3a$_1^{-1}$ & 1b$_1^{-1}$ & Total \\ \midrule
2a$_1$2a$_1$ & 8.9 & -- & -- & -- & 8.9 & 9.1 \\
2a$_1$3a$_1$ (singlet) & 4.8 & -- & 2.4 & -- & 7.2 & 6.8 \\
2a$_1$3a$_1$ (triplet) & 1.0 & -- & 1.7 & -- & 1.4 & 1.3 \\
2a$_1$1b$_1$ (singlet) & 4.5 & -- & -- & 1.9 & 6.4 & 6.5 \\
2a$_1$1b$_1$ (triplet) & 1.2 & -- & -- & 1.9 & 1.5 & 1.6 \\
2a$_1$1b$_2$ (singlet) & 2.9 & 1.6 & -- & -- & 4.5 & 4.1 \\
2a$_1$1b$_2$ (triplet) & 0.7 & 1.6 & -- & -- & 1.2 & 1.0 \\
3a$_1$3a$_1$ & -- & -- & 5.7 & -- & 5.7 & 5.9 \\
3a$_1$1b$_1$ (singlet) & -- & -- & 4.2 & 3.7 & 7.9 & 8.3 \\
3a$_1$1b$_1$ (triplet) & -- & -- & 0.1 & 0.2 & 0.2 & 0.1 \\
3a$_1$1b$_2$ (singlet) & -- & 3.3 & 3.3 & -- & 6.6 & 8.3 \\
3a$_1$1b$_2$ (triplet) & -- & 0.2 & 0.1 & -- & 0.1 & 0.1 \\
1b$_1$1b$_1$ & -- & -- & -- & 7.6 & 7.6 & 8.1 \\
1b$_1$1b$_2$ (singlet) & -- & 3.7 & -- & 3.2 & 6.9 & 6.9 \\
1b$_1$1b$_2$ (triplet) & -- & 0.0 & -- & 0.0 & 0.0 & 0.0 \\
1b$_2$1b$_2$ & -- & 4.3 & -- & -- & 4.3 & 5.0 \\ \midrule
Sum & 24.0 & 14.5 & 18.6 & 17.5 & 70.5 & 73.1 \\ \midrule
$-$Im($E)$ & 24.8 & 16.2 & 18.1 & 18.4 & 77.5 & 70.X \\ \bottomrule
\end{tabular}\hspace{1.5cm}}
\label{tab:h2opwcis}
\end{table}

The double counting of triplet channels is avoided by computing partial decay 
widths using ACPs and then summing over the singlet channels while averaging 
the values for the triplet channels as discussed in Sec.~\ref{sec:cistheory}. 
In Table~\ref{tab:h2opwcis}, we report the partial widths obtained with 
``out''-ACPs, while results computed with ``in''-ACPs are available in 
the Supplementary Material. Our results demonstrate that ACP-CIS produces 
partial widths in remarkable agreement with those from the decomposition 
of the CCSD energy, the RMS deviation amounts to only 0.5 meV. The largest 
deviation between ACP-CIS and CCSD (1.7 meV) occurs for the 3a$_1$1b$_2$ 
singlet channel, for all other channels it stays below 1 meV. Recombining 
the partial widths to a total decay half-width results in 70.5 meV, which 
is very close to the numbers obtained with $\Delta$CCSD (70~meV) and 
EOMIP-CCSD (71~meV).

\begin{figure}
\centering
\includegraphics[width=\linewidth]{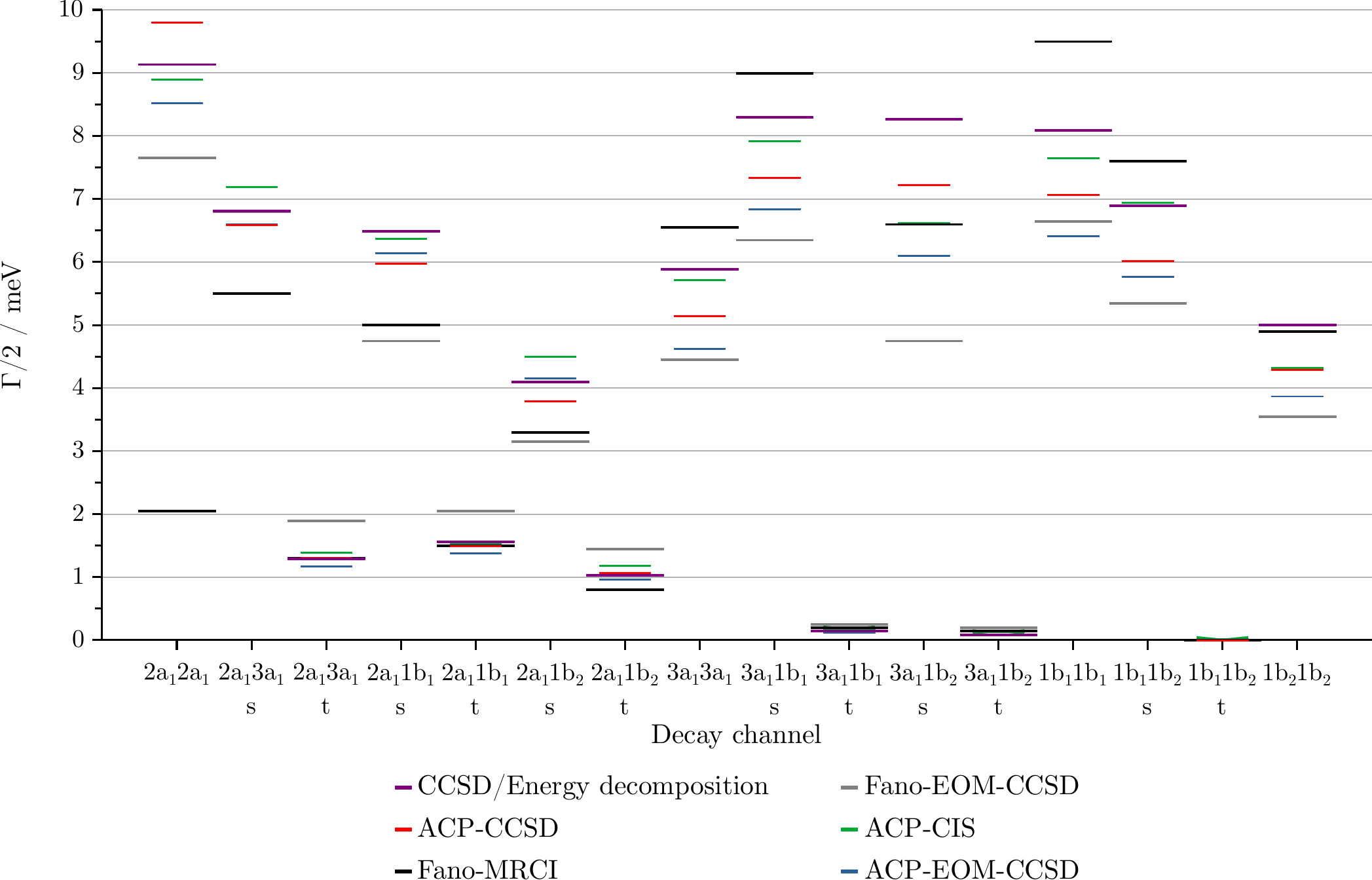}
\caption{Partial decay widths of core-ionized water computed with different 
methods. Complex-variable calculations are done with the cc-pCVTZ (5sp) basis 
set with 2 complex scaled s-, p- and d-shells on each atom; all values are 
available from Tables \ref{tab:h2opw} and \ref{tab:h2opwcis}. Fano-MRCI and 
Fano-EOM-CCSD results are taken from Refs. \cite{inhester12,inhester14} and 
\cite{skomorowski21b}, respectively.}
\label{fig:h2opw}
\end{figure}

This good agreement of CIS with CCSD and EOMIP-CCSD demonstrates that 
the individual decay channels are, in general, well described by CIS. 
The representation of a channel by a single configuration according to 
Figure~\ref{fig:cis} appears to be a valid approximation. However, there 
are a few decay channels where this may not be the case. This is apparent 
from Figure~\ref{fig:h2opw} where results from different methods for the 
partial widths of core-ionized water are compared. Even though the overall 
agreement between all methods is good, there are a few conspicuous cases 
where Fano-MRCI deviates from all other results, in particular 2a$_1$2a$_1$ 
and to a lesser degree 2a$_1$3a$_1$ and 1b$_1$1b$_1$. This may indicate 
that these channels are not so well described by a single configuration, 
which may be captured better by MRCI.

\section{Further applications: Ne, NH$_3$, and CH$_4$}\label{sec:appl}

As further applications of ACPs, we investigated Auger decay of core-ionized states of 
methane (CH$_4$), ammonia (NH$_3$) and the neon atom. Like H$_2$O, all these species 
have 10 electrons in their neutral ground state. The same basis sets as in Sec. \ref{sec:h2o} 
are used, details are available from the Supplementary Material. 

\subsection{Total decay widths} \label{sec:totwidth}

Our results for total half-widths are summarized in Table~\ref{tab:netw}. 
For CIS, ``out''-ACPs and a uniform $\theta_\mathrm{opt}$ were used as discussed in Sec. 
\ref{sec:cisresults}. Additional results including $\theta_\mathrm{opt}$ values, 
CIS results with ``in''-ACPs, and CIS contributions from individual calculations 
can be found in the Supplementary Material. 

The different methods yield very similar decay half-widths for all species: no deviation
larger than 5 meV is observed. Our results are also in qualitative agreement with 
experimental values~\cite{muller17,sankari03,carroll99,asplund85,heimann91,koeppe96} 
and with decay widths computed using Fano's theory~\cite{skomorowski21b}. In particular, 
EOMIP-CCSD, $\Delta$CCSD, and CIS all produce the correct dependence of the Auger decay 
width on the nuclear charge $Z$, that is, higher $Z$ leads to faster decay. This trend 
is well-known and has been related to the higher kinetic energy of the Auger electron 
due to higher $Z$~\cite{agarwal13,mcguire69,hasoglu08}.

We add that there is disagreement between different experiments about the Auger decay 
width of methane~\cite{carroll99,asplund85,heimann91,koeppe96} with values for the 
half-width ranging from 41 to 60~meV. Our results as well as Fano's theory are more 
in line with the lower values, but the results for the other species also suggest that 
our methods and Fano's theory both may underestimate the decay width systematically. 
This disagreement with experimental values may be caused by processes involving more 
than two electrons, which are not described by any of our methods. However, it may 
also be ascribed in part to the basis set. Notably, we employed in our previous 
work~\cite{matz22} the ``aug-cc-pCVTZ (5sp) + 2$\times$(spd)'' basis set (instead 
of ``cc-pCVTZ (5sp) + 2$\times$(spd)'' here) and obtained values of 75 and 78 meV 
for the total half-width of H$_2$O using EOMIP-CCSD and $\Delta$CCSD, respectively. 
Also, while 2 complex-scaled s, p, and d-shells are sufficient for H$_2$O, 4 of them 
appear to be necessary for the neon atom. These differences are documented in the 
Supplementary Material. 

\begin{table}[h]
\tbl{Total decay half-widths in meV of the core-ionized neon atom and the water, ammonia, 
and methane molecules.}
{\hspace{3cm}\begin{tabular}{lcccc}\toprule
Method & Ne & H$_2$O & NH$_3$ & CH$_4$ \\ \midrule 
EOMIP-CCSD$^a$ & 110 & 71 & 56 & 37 \\
$\Delta$CCSD$^a$ & 115 & 70 & 55 & 38 \\
CIS$^b$ & 113 & 71 & 53 & 37 \\ \midrule
Fano-EOM-CCSD$^c$ & 109 & 61 & -- & 33 \\
Experiment$^d$ & 129 & 80 & -- & 47 \\ \bottomrule
\end{tabular}\hspace{3cm}}
\tabnote{$^a$ This work. Computed with the cc-pCVTZ (5sp) + 2$\times$(spd) basis set 
[Ne: 4$\times$(spd)] and evaluated as imaginary part of the total energy.}
\tabnote{$^b$ This work. Computed with the cc-pCVTZ (5sp) + 2$\times$(spd) basis set 
[Ne: 4$\times$(spd)] and ``out''-ACPs, evaluated as weighted sum of partial widths, 
see Sec. \ref{sec:cistheory} for details.}
\tabnote{$^c$ From Ref. \cite{skomorowski21b}. Computed using Coulomb waves for the 
Auger electron. For H$_2$O and CH$_4$, values of $Z_\mathrm{eff}$ of 4.9 (H$_2$O) 
and 3.6 (CH$_4$) were assumed for all channels, while for Ne, different values were 
assumed for each channel.}
\tabnote{$^d$ From Refs. \cite{muller17} (Ne), \cite{sankari03} (H$_2$O), 
and \cite{carroll99} (CH$_4$).}
\label{tab:netw}
\end{table}

%%%%%%%%%%%%%%%%%%%%%%%%%%%%%%%%%%%%%%%%%%%%%%%%%%%%%%%%%%%%%%%%%%%%%%%%%%%%%%%%%%%%%%%%%%

%To illustrate the agreement between different methods, the root mean square deviations between the partial widths calculated with each method and those obtained via decomposition of the CCSD energy after convergence were determined and are compared in figure~\ref{fig:rmsacp}.

% For each molecule, the different methods result in similar optimal scaling angles and
% close agreements between the total decay widths, except for the neon atom, where the
% results for the total decay widths differ somewhat from each other. An inaccuracy of the
% results for neon may also be indicated by higher values of $\mathrm{d}E/\mathrm{d}\theta$
% compared to other systems. 

%However, in each method one still obtains the accurate result
%for the dependency of the decay widths on the nuclear charge: the higher it is, the faster
%the decay process. This dependency of the Auger transition probability on the kinetic
%energy of the Auger electron and therefore on the nuclear charge is
%well-known~\cite{agarwal13,hasoglu08}.

%%%%%%%%%%%%%%%%%%%%%%%%%%%%%%%%%%%%%%%%%%%%%%%%%%%%%%%%%%%%%%%%%%%%%%%%%%%%%%%%%%%%%%%%%

\subsection{Partial decay widths} \label{sec:pwidth}

Partial decay half-widths for Ne, NH$_3$, and CH$_4$ are available from Tables~\ref{tab:nepw}, 
\ref{tab:nh3pw}, and~\ref{tab:ch4pw}, respectively. Further results and a graphical 
comparison of different methods are available from the Supplementary Material. In 
line with previous results~\cite{matz22}, we observe that partial widths and branching 
ratios are significantly more sensitive towards the computational details than total 
widths. Details are discussed in the following. 

\begin{table} [h]
\tbl{Partial decay half-widths of core-ionized neon in meV.}
{\begin{tabular}{lrrrrrr}\toprule
Decay & CCSD & EOMIP-CCSD & CIS & CCSD & EOM-CCSD & \\ 
channel & \multicolumn{3}{c}{out-ACP} & energy decomp. & Fano\textsuperscript{a} & 
expt.\textsuperscript{b} \\ \midrule 
$^1$D (2p$^{-2}$) & 69.4 & 63.7 & 68.0 & 75.5 & 58.8 & 78.2(21) \\%78
$^1$P (2s$^{-1}$2p$^{-1}$) & 28.1 & 25.8 & 23.0 & 29.6 & 19.6 & 22.1(7) \\%22
$^3$P (2s$^{-1}$2p$^{-1}$) & 6.9 & 5.6 & 5.4 & 6.7 & 11.9 & 8.1(3) \\%8
$^1$S (2p$^{-2}$) & 4.9 & 4.5 & 3.1 & 5.7 & 13.6 & 7.9(3) \\%8
$^1$S (2s$^{-2}$) & 13.8 & 12.3 & 13.3 & 13.5 & 5.3 & 12.2(4) \\ \bottomrule
\end{tabular}}
\tabnote{\textsuperscript{a} From reference~\cite{skomorowski21b}.\\
\textsuperscript{b} From reference~\cite{muller17}.}
\label{tab:nepw}
\end{table}

The core-ionized neon atom with the electronic configuration of (1s)$^1$(2s)$^2$(2p)$^6$ 
has 5 main decay channels. Their relative importance is captured correctly by all methods: 
the $^1$D channel is broadest followed by the $^1$P channel. Table \ref{tab:nepw} shows 
that the ACP methods tend to produce systematically smaller widths for all channels than 
the decomposition of the CCSD energy. This tendency is also present for H$_2$O (Sec. 
\ref{sec:acpresults}, Tab.~\ref{tab:h2opw}), but it is a lot more pronounced for the 
$^1$D and $^1$P channels of Ne. For the $^1$D channel, it leads to the ACP methods 
underestimating the experimental width by 10--20\%, while CCSD energy decomposition 
agrees within 3\% with the experiment. However, for the next-broadest channel ($^1$P) 
the performance is reversed and ACP-CIS is closest to the experimental value. Notably, 
the $^1$D and $^1$P channels are also most sensitive to the complex-scaled basis set. 
As shown in the Supplementary Material, the respective branching ratio is significantly 
distorted when only 2 instead of 4 complex-scaled shells are employed, while all 
remaining channels, which are narrower, are less sensitive. 

\begin{table}[h]
\tbl{Partial decay half-widths of core-ionized ammonia in meV.}
{\begin{tabular}{lrrrr}\toprule
Decay & CCSD & EOMIP-CCSD & CIS & CCSD \\ 
channel & \multicolumn{3}{c}{out-ACP} & energy decomp. \\ \midrule
$^1$A$_1$ (2a$_1^{-2}$) & 7.8 & 7.2 & 7.6 & 7.3 \\
$^1$E (2a$_1^{-1}$1e$^{-1}$) & 7.2 & 7.7 & 7.7 & 7.9 \\
$^3$E (2a$_1^{-1}$1e$^{-1}$) & 2.0 & 1.7 & 2.1 & 2.0 \\
$^1$A$_1$ (1e$^{-2}$) & 2.6 & 2.5 & 1.9 & 3.2 \\
$^1$E (1e$^{-2}$) & 10.2 & 8.3 & 9.5 & 12.1 \\
$^3$A$_2$ (1e$^{-2}$) & 0.0 & 0.0 & 0.0 & 0.0 \\
$^1$A$_1$ (3a$_1^{-2}$) & 5.8 & 4.8 & 5.8 & 6.7 \\
$^1$A$_1$ (2a$_1^{-1}$3a$_1^{-1}$) & 5.2 & 5.3 & 5.6 & 5.3 \\
$^3$A$_1$ (2a$_1^{-1}$3a$_1^{-1}$) & 1.1 & 1.0 & 1.1 & 1.1 \\
$^1$E (3a$_1^{-1}$1e$^{-1}$) & 12.2 & 10.3 & 11.5 & 14.2 \\
$^3$E (3a$_1^{-1}$1e$^{-1}$) & 0.1 & 0.1 & 0.2 & 0.1 \\ \bottomrule
\end{tabular}}
\label{tab:nh3pw}
\end{table}

%12.05.2022: Feshbach-Fano values now accurate
\begin{table}
\tbl{Partial decay half-widths of core-ionized methane in meV.}
{\begin{tabular}{lrrrrr}\toprule
Decay & CCSD & EOMIP-CCSD & CIS & CCSD & EOM-CCSD \\ 
channel & \multicolumn{3}{c}{out-ACP} & energy decomp. & Fano\textsuperscript{a} \\ \midrule
$^1$A$_1$ (2a$_1^{-2}$) & 5.8 & 5.9 & 5.7 & 5.9 & 6.6 \\
$^1$T$_2$ (2a$_1^{-1}$1t$_2^{-1}$) & 8.7 & 7.1 & 9.5 & 8.6 & 7.1 \\
$^3$T$_2$ (2a$_1^{-1}$1t$_2^{-1}$) & 2.3 & 2.4 & 2.9 & 2.9 & 3.4 \\
$^1$T$_2$ (1t$_2^{-2}$) & 12.0 & 14.0 & 19.3 & 16.3 & 8.0 \\
$^1$A$_1$ (1t$_2^{-2}$) & 1.1 & 1.1 & 1.3 & 0.9 & 3.0 \\
$^1$E (1t$_2^{-2}$) & 6.5 & 6.3 & 9.9 & 8.3 & 5.4 \\
$^3$T$_1$ (1t$_2^{-2}$) & 0.0 & 0.1 & 0.0 & 0.0 & 0.0 \\ \bottomrule
\end{tabular}}
\tabnote{\textsuperscript{a} From Ref.~\cite{skomorowski21b}.}
\label{tab:ch4pw}
\end{table}

The core-ionized ammonia molecule with the electronic configuration 
(1a$_1$)$^1$(2a$_1$)$^2$(1e)$^4$(3a$_1$)$^2$ has 11 main decay channels. As 
Table \ref{tab:nh3pw} illustrates, all our methods agree that two $^1$E channels 
(1e$^{-2}$ and 3a$_1^{-1}$1e$^{-1}$) are broadest followed by a third $^1$E 
channel (2a$_1^{-1}$1e$^{-1}$) and a $^1$A$_1$ channel (2a$_1^{-2}$). This 
overall consistency between the methods is also reflected by the good agreement 
in the total width (Tab.~\ref{tab:netw}). Especially remarkable is the good 
performance of CIS because these calculations are based on symmetry-broken HF 
determinants and the identification of the decay channels is dubious in this 
case as we discussed in Sec. \ref{sec:symtheory}. From Tab. \ref{tab:nh3pw}, it 
is also apparent that, similar to neon, all ACP methods yield smaller widths 
than energy decomposition on average. This is most pronounced for the two 
broadest $^1$E channels. Together with the fact that EOMIP-CCSD yields somewhat 
smaller widths than CCSD for some channels, significantly different branching 
ratios can be obtained from the different methods in some cases. 

The core-ionized methane molecule with the electronic configuration 
(1a$_1$)$^1$(2a$_1$)$^2$(2t$_2$)$^6$ features 7 decay channels. Table \ref{tab:ch4pw} 
shows that all our methods as well as Fano-EOM-CCSD agree that the $^1$T$_2$ (1t$_2^{-2}$) 
channel is broadest but apart from this there is substantial disagreement between the 
methods about most singlet channels. While there are some channels with small deviations, 
in particular $^1$A$_1$ (2a$_1^{-2}$), $^1$T$_2$ (2a$_1^{-1}$1t$_2^{-1}$), and $^1$A$_1$ 
(1t$_2^{-2}$), the widths of the $^1$E channels and the other $^1$T$_2$ channels differ 
substantially between the methods. On average, decomposition of the CCSD energy yields 
the largest partial widths for CH$_4$, whereas Fano-EOM-CCSD yields the smallest widths 
with the ACP results falling in between. The same trend as for H$_2$O, Ne, and NH$_3$ 
is thus observed; it is, however, more pronounced for CH$_4$ and Ne than for NH$_3$ 
and H$_2$O. 

Taking a step back and comparing the different species (H$_2$O, NH$_3$, CH$_4$ and Ne), 
we observe that the agreement between ACPs, energy decomposition, and Fano's theory is 
best for totally symmetric decay channels, whereas substantial differences occur for 
other channels, especially for decay into degenerate final states. For H$_2$O, there 
are no degenerate states so that the agreement is much better than for the other 
species. A detailed investigation of this phenomenon is beyond the present work, 
but based on the good agreement with experimental values for neon, we think that the 
energy decomposition approach gives the best results. This would mean that removing 
certain decay channels from the excitation manifold as done in the ``out''-ACP approach 
works better if these channels are totally symmetric. We reiterate the different 
basis-set dependence of H$_2$O and Ne in this context: While using 2 instead of 4 
complex-scaled shells makes little impact on all H$_2$O results, the width of the 
$^1$D channel of Ne changes considerably.

\subsection{Inner-valence and outer-valence orbitals}

While it is well known that the total Auger decay width of a core hole increases with 
nuclear charge $Z$~\cite{agarwal13,mcguire69,hasoglu08}, trends in partial widths are more involved. 
To compare partial widths between CH$_4$, NH$_3$, H$_2$O, and Ne, we distinguish 
inner-valence and outer-valence orbitals: The 2a$_1$ orbitals in water, ammonia, 
and methane and the 2s orbital in neon are defined as inner-valence, while all other 
higher-lying orbitals are defined as outer-valence. This distinction is useful because 
holes in inner-valence and outer-valence orbitals relax in different ways. In particular, 
only inner-valence holes can undergo ICD~\cite{cederbaum97,jahnke20}, whereas outer-valence 
holes do not have enough energy. Partial Auger decay widths involving inner-valence orbitals 
thus influence ICD rates.

\begin{figure}[tbh]
\centering
\includegraphics[width=\linewidth]{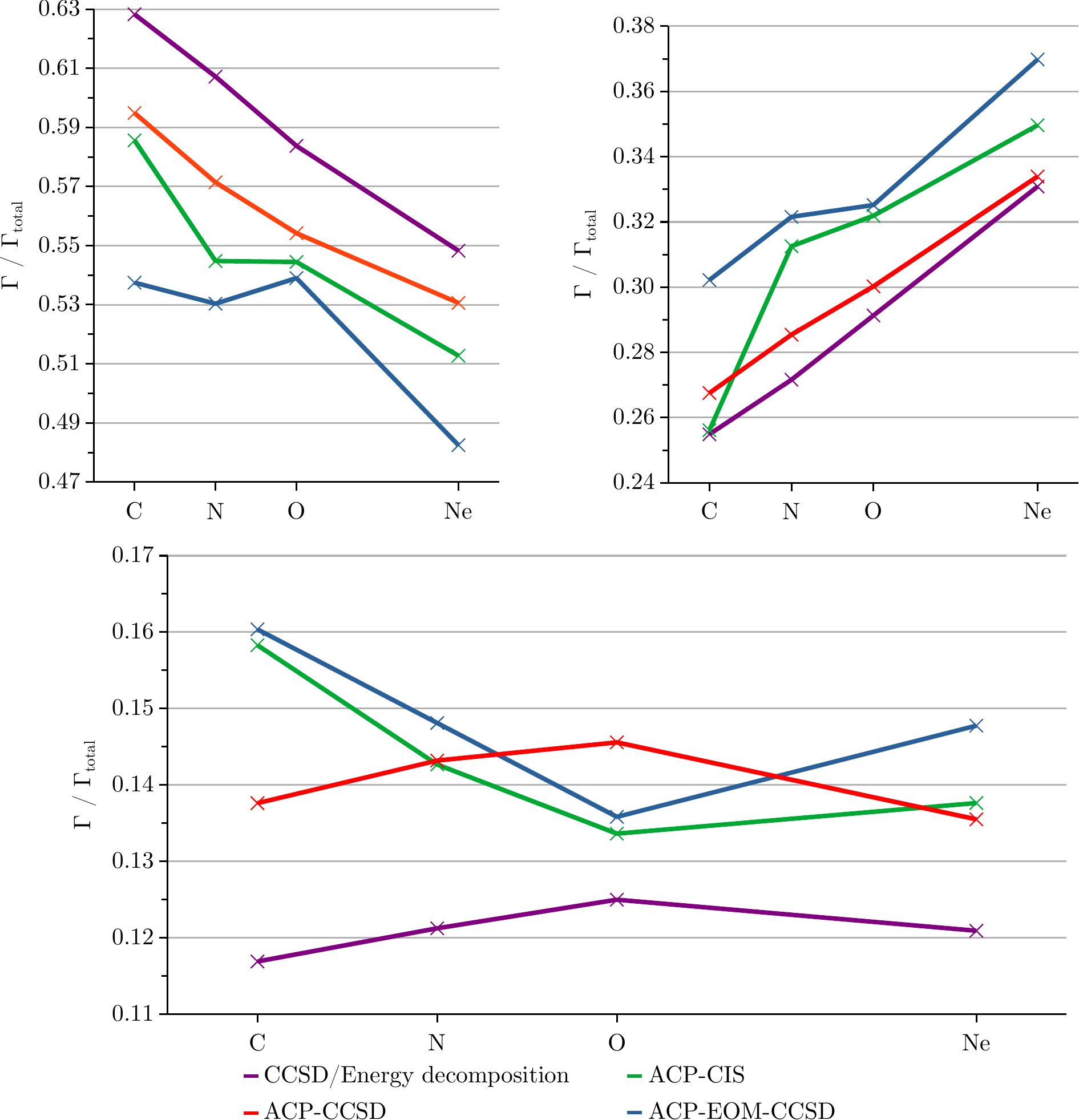}
\caption{Relative contributions of decay channels involving two outer-valence spin-orbitals 
(upper left), two inner-valence spin-orbitals (bottom), and one outer-valence and one 
inner-valence spin-orbital (upper right) to the total Auger decay width of CH$_4$, NH$_3$, 
H$_2$O, and Ne.}
\label{fig:all-trends}
\end{figure}
 
For all systems investigated here, there are 2 inner-valence spin-orbitals and 6 outer-valence 
spin-orbitals. From statistical considerations, one would thus expect the pure inner-valence 
channel to contribute $1/16 \approx 6\%$ to the total decay width. Figure~\ref{fig:all-trends} 
shows that this is not the case, the inner-valence decay channel rather contributes 12--16\% 
depending on the system and method. Furthermore, Figure~\ref{fig:all-trends} demonstrates 
that the relative contribution of decay channels involving two outer-valence spin-orbitals 
decreases with growing $Z$. In return, decay involving one inner-valence and one outer-valence 
spin-orbital becomes more important, while the contribution of the pure inner-valence channel 
depends only weakly on $Z$. All our methods agree on these trends, although there is a minor 
disagreement about the channel involving two inner-valence spin-orbitals: CIS and EOMIP-CCSD 
find the highest relative contribution of this channel for H$_2$O where CCSD finds the lowest 
contribution. In sum, our results indicate that ICD and related processes that require 
inner-valence holes are more likely to occur after Auger decay in neon than in molecules 
with lighter nuclei.

\section{Conclusions and Outlook} \label{sec:conc}

We have introduced Auger channel projectors (ACPs) for the determination of partial 
Auger decay widths from complex-variable electronic-structure calculations. ACPs 
are related to the projectors used for implementing the core-valence separation but 
specific to particular decay channels. 

While we showed previously that partial Auger decay widths can be computed by 
decomposing the imaginary part of the CCSD energy~\cite{matz22}, the ACP approach 
can be combined with EOMIP-CCSD and CIS as well. Results from both approaches 
are generally in good agreement although widths for decay into degenerate states 
of molecules with high symmetry appear to be systematically underestimated by ACP 
methods. A further advantage of the energy decomposition approach is that it is 
always based on a single CCSD calculation irrespective of the number of decay 
channels, whereas the ACP approaches require to determine the wave function anew 
for each decay channel. On the other hand, because EOMIP-CCSD is less costly than 
CCSD, ACP-EOMIP-CCSD is potentially computationally cheaper than CCSD energy 
decomposition. Moreover, the EOMIP-CCSD wave function is a spin eigenfunction 
and there is no need to solve the HF and CCSD equations for the core-vacant state, 
which can be problematic. 

Also, we showed that accurate total and partial Auger decay widths can be computed 
with ACPs applied to CIS wave functions although an individual CIS wave function 
describes only a subset of the decay channels. While CIS is attractive in terms 
of computational cost, the main drawback of its application to Auger decay is that 
this requires to determine the HF wave functions of all states with a hole in a 
valence shell. This may be impractical for larger molecules and may also be affected 
by convergence problems. 

We conclude by remarking that ACPs are not only useful for modeling Auger decay, but 
can be applied to resonant Auger decay as well as decay processes that do not involve 
core orbitals such as ICD.

\section*{Acknowledgements}

F.M. thanks Garrette Pauley Paran, Anthuan Ferino Pérez, Valentina Parravicini, and Jonas 
Nijssen for helpful discussions about point-group symmetry and for help in verifying the 
implementation. T.-C. J. gratefully acknowledges funding from the European Research Council 
(ERC) under the European Union's Horizon 2020 research and innovation program (Grant Agreement 
No. 851766). F. M. is grateful for a Kekulé fellowship (K 208/24) by the Fonds der Chemischen
Industrie.

\end{document}

\maketitle
\centering

\captionof{table}{Exponents used for complex-scaled basis functions in CBF calculations.}
\begin{tabular}{lS[table-format=1.6,round-mode=places,round-precision=6]S[table-format=2.6,
round-mode=places,round-precision=6]S[table-format=2.6,round-mode=places,round-precision=6]}\toprule
    Atom & {s} & {p} & {d}\\\midrule
    \multicolumn{4}{c}{Basis set: cc-pCVTZ (5sp) + 2$\times$(spd)}\\\midrule
    \multirow{2}{*}{H}&0.3988926&1.6055999&2.1842133\\
    &0.1292174&0.5205513&0.3861180\\\midrule
    \multirow{2}{*}{C}&1.5775008&6.3496660&8.6379090\\
    &0.5110162&2.0586245&1.5269810\\\midrule
    \multirow{2}{*}{N}&2.1804898&8.7767831&11.9396915\\
    &0.7063487&2.8455199&2.1106592\\\midrule
    \multirow{2}{*}{O}&2.7759139&11.1734500&15.2000506\\
    &0.8992306&3.6225430&2.6870147\\\midrule
    \multirow{2}{*}{Ne}&4.3306000&17.4312839&23.7130337\\
    &1.4028562&5.6513946&4.1919117\\\midrule
    \multicolumn{4}{c}{Basis set: cc-pCVTZ (5sp) + 4$\times$(spd)}\\\midrule
    \multirow{4}{*}{H}&0.3988926&1.6055999&2.1842133\\
    &0.2584348&1.0411026&0.7722360\\
    &0.1292174&0.5205513&0.3861180\\
    &0.0646087&0.2602757&0.1930590\\\midrule
    \multirow{4}{*}{O}&2.7759139&11.1734500&15.2000506\\
    &1.7984612&7.2450860&5.3740294\\
    &0.8992306&3.6225430&2.6870147\\
    &0.4496153&1.8112715&1.3435074\\\midrule
    \multirow{4}{*}{Ne}&4.3306000&17.4312839&23.7130337\\
    &2.8057124&11.3027892&8.3838234\\
    &1.4028562&5.6513946&4.1919117\\
    &0.7014281&2.8256973&2.0959559\\\bottomrule
\end{tabular}\\[1ex]

%%%%%%%%%%%%%%%%%%%%%%%%%%%%%%%%%%%%%%%%%%%%%%%%%%%%%%%%%%%%%%%%%%%%%%%%%%%%%%%%%%%
\newpage
\captionof{table}{Results for the core-ionized state of water from CIS calculations 
using different states of the water cation as HF reference. The cc-pCVTZ (5sp) basis set 
with 2 complex scaled s-, p- and d-shells on each atom is used. $\Delta E$ refers to the 
energy difference between valence-hole reference state and core-ionized state. Calculations 
were carried out with a $\theta$ optimized for each reference (upper part of the table) 
and with a uniform $\theta$ (lower part of the table). The last row shows the sum of the 
partial half-widths of the channels included in the respective CIS calculation as obtained 
from a decomposition of the CCSD energy. Here, only 50\% of the width is considered for 
decay into open-shell singlet states as only 50\% of these wave functions is described by CIS.}
\begin{tabular}{lllll}\toprule
HF reference state & 2a$_1^{-1}$ & 1b$_2^{-1}$ & 3a$_1^{-1}$ & 1b$_1^{-1}$ \\ \midrule
Decay channels & 2a$_1$2a$_1$, 2a$_1$3a$_1$, & 2a$_1$1b$_2$, 3a$_1$1b$_2$, & 
2a$_1$3a$_1$, 3a$_1$3a$_1$, & 2a$_1$1b$_1$, 3a$_1$1b$_1$, \\
included & 2a$_1$1b$_1$, 2a$_1$1b$_2$ & 1b$_1$1b$_2$, 1b$_2$1b$_2$ & 
3a$_1$1b$_1$, 3a$_1$1b$_2$ & 1b$_1$1b$_1$, 1b$_1$1b$_2$ \\ \midrule 
$\theta_\mathrm{opt}$ / $^\circ$ & 22 & 28 & 28 & 14 \\
$|\mathrm{d}E/\mathrm{d}\theta|$ / Hartree/$^\circ$ & $4.5\cdot 10^{-6}$ & 
$1.2\cdot 10^{-5}$ & $1.7\cdot 10^{-5}$ & $2.4\cdot 10^{-5}$ \\
$\mathrm{Re}(\Delta E)$ / eV & 514.0 & 530.0 & 533.4 & 535.1 \\
$-\mathrm{Im}(\Delta E)$ / meV & 19.6 & 11.1 & 20.1 & 14.4 \\
$-\mathrm{Im}(E)$ / meV & 26.6 & 16.2 & 18.4 & 17.1 \\ \midrule
$\theta$ / $^\circ$ & 28 & 28 & 28 & 28 \\
$|\mathrm{d}E/\mathrm{d}\theta|$ / Hartree/$^\circ$ & $9.9\cdot 10^{-6}$ & 
$1.2\cdot 10^{-5}$ & $1.7\cdot 10^{-5}$ & $4.3\cdot 10^{-5}$ \\
$\mathrm{Re}(\Delta E)$ / eV & 514.0 & 530.0 & 533.4 & 535.1 \\
$-\mathrm{Im}(\Delta E)$ / meV & 20.0 & 11.1 & 20.1 & 13.6 \\
$-\mathrm{Im}(E)$ / meV & 24.8 & 16.2 & 18.4 & 18.1 \\ \midrule
$\sum\Gamma_\mathrm{CCSD}/2$ / meV & 21.7 & 15.8 & 19.1 & 20.6 \\ \bottomrule
\end{tabular}

%%%%%%%%%%%%%%%%%%%%%%%%%%%%%%%%%%%%%%%%%%%%%%%%%%%%%%%%%%%%%%%%%%%%%%%%%%%%%%%%%%
\vspace{1cm}

\captionof{table}{Partial half-widths of core-ionized water, calculated with different methods 
and the cc-pCVTZ (5sp) basis set with 2 complex scaled s-, p- and d-shells on each atom.}
\begin{tabular}{lllll|rrrrr}\toprule
 & \multicolumn{4}{c|}{CIS} & CCSD & EOMIP-CCSD & \multicolumn{2}{c}{CCSD} & MRCI \\
Decay & \multicolumn{2}{c}{ACP/id. $\theta$} & \multicolumn{2}{c|}{ACP/diff. $\theta$} 
& ACP & ACP & \multicolumn{2}{c}{energy} & Fano\textsuperscript{b} \\
channel & ``in'' & ``out'' & ``in'' & ``out'' & ``out'' & ``out'' & \multicolumn{2}{c}{decomp.\textsuperscript{a}} & \\\midrule
2a$_1$2a$_1$ & 8.7 & 8.8 & 8.8 & 9.1 & 9.8 & 8.5 & 9.1 & 9.7 & 2.1 \\
2a$_1$3a$_1$ (singlet) & 7.0 & 7.2 & 6.8 & 7.1 & 6.6 & 6.6 & 6.8 & 8.6 & 5.5 \\
2a$_1$3a$_1$ (triplet) & 1.2 & 1.4 & 1.2 & 1.4 & 1.3 & 1.2 & 1.3 & 1.8 & 1.3 \\
2a$_1$1b$_1$ (singlet) & 6.2 & 6.4 & 6.3 & 6.5 & 6.0 & 6.1 & 6.5 & 7.9 & 5.0 \\
2a$_1$1b$_1$ (triplet) & 1.4 & 1.5 & 1.4 & 1.6 & 1.5 & 1.4 & 1.6 & 1.9 & 1.5 \\
2a$_1$1b$_2$ (singlet) & 4.4 & 4.5 & 4.1 & 4.2 & 3.8 & 4.2 & 4.1 & 6.2 & 3.3 \\
2a$_1$1b$_2$ (triplet) & 1.1 & 1.2 & 1.0 & 1.2 & 1.1 & 1.0 & 1.0 & 1.5 & 0.8 \\
3a$_1$3a$_1$ & 5.6 & 5.7 & 5.6 & 5.7 & 5.1 & 4.6 & 5.9 & 7.3 & 6.6 \\
3a$_1$1b$_1$ (singlet) & 7.8 & 7.9 & 7.1 & 7.2 & 7.3 & 6.9 & 8.3 & 10.9 & 9.0 \\
3a$_1$1b$_1$ (triplet) & 0.2 & 0.2 & 0.2 & 0.2& 0.1 & 0.1 & 0.1 & 0.1 & 0.2 \\
3a$_1$1b$_2$ (singlet) & 6.5 & 6.6 & 6.5 & 6.6 & 7.2 & 6.1 & 8.3 & 8.3 & 6.6 \\
3a$_1$1b$_2$ (triplet) & 0.1 & 0.1 & 0.1 & 0.1 & 0.1 & 0.1 & 0.1 & 0.1 & 0.2 \\
1b$_1$1b$_1$ & 7.6 & 7.6 & 6.7 & 6.6 & 7.1 & 6.4 & 8.1 & 9.8 & 9.5 \\
1b$_1$1b$_2$ (singlet) & 6.8 & 6.9 & 6.1 & 6.2 & 6.0 & 5.8 & 6.9 & 9.1 & 7.6 \\
1b$_1$1b$_2$ (triplet) & 0.0 & 0.0 & 0.0 & 0.0 & 0.0 & 0.0 & 0.0 & 0.0 & 0.0 \\
1b$_2$1b$_2$ & 4.3 & 4.3 & 4.3 & 4.3 & 4.3 & 3.9 & 5.0 & 5.4 & 4.9 \\ \midrule
Sum & 68.8 & 70.5 & 66.3 & 67.9 & 67.3 & 62.8 & 73.1 & 88.5 & 64.1 \\ \bottomrule
\end{tabular}\\[1ex]
\begin{raggedright}
\textsuperscript{a} Right column: Obtained with the cc-pCVTZ (5sp) basis set with 4 complex scaled s-, p- and d-shells on each atom.\\[1ex]
\textsuperscript{b} From Refs. 24 and 55.\\
\end{raggedright}

\vspace{1cm}

\includegraphics[width=\linewidth]{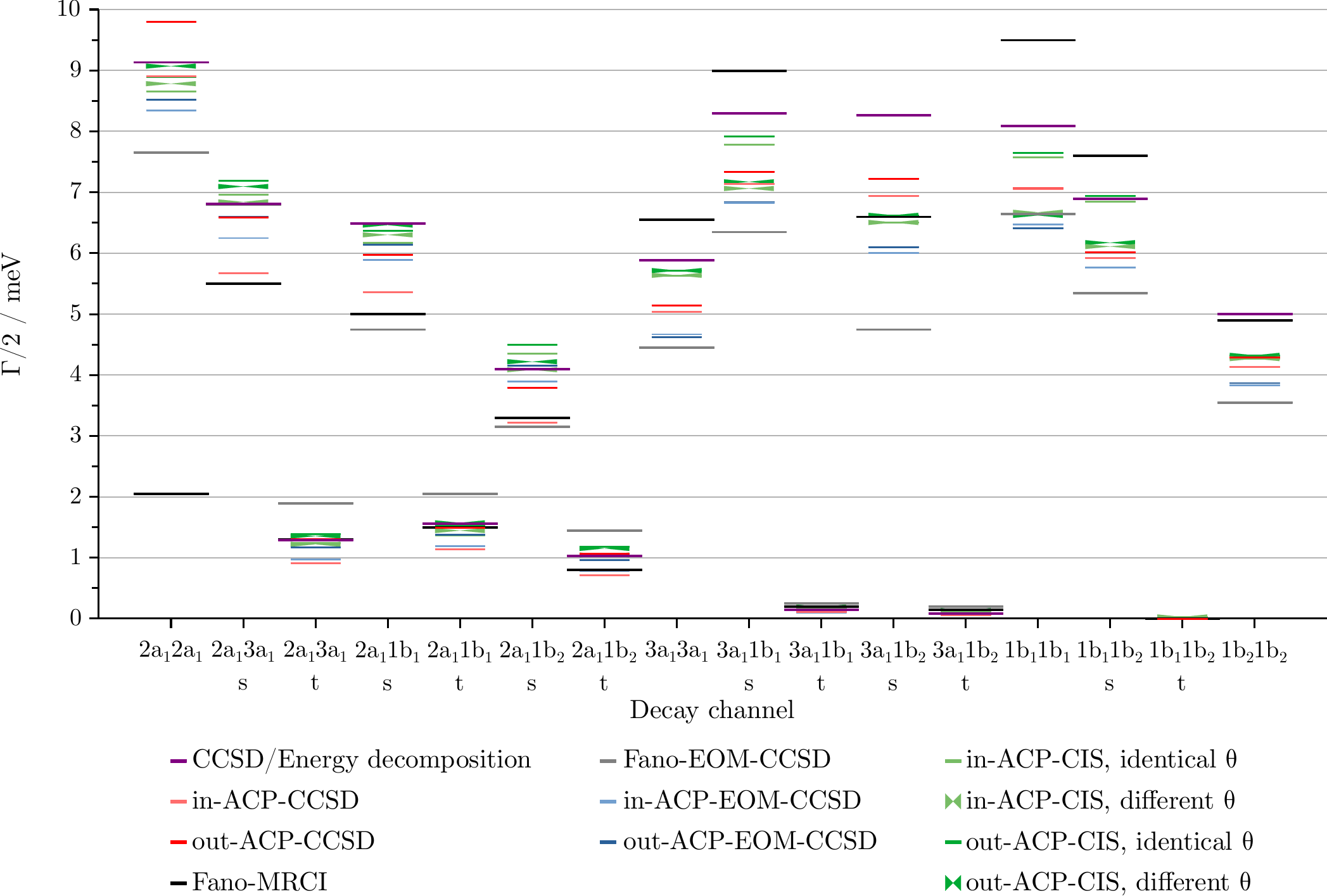}
\captionof{figure}{Partial decay widths of core-ionized water computed with different 
methods. Complex-variable calculations are done with the cc-pCVTZ (5sp) basis set with 
2 complex scaled s-, p- and d-shells on each atom; all values are available from Table 3. 
Fano-MRCI and Fano-EOM-CCSD results are taken from Refs. [24,55] and [26], respectively.}

%\includegraphics[width=\linewidth]{RMS-acp.pdf}
%\captionof{figure}{Root mean square deviation between the partial widths calculated with each method and the results of CCSD energy decomposition, normalized by the average partial width in the CCSD energy decomposition, each calculated with the cc-pCVTZ (5sp) basis set with 2 complex scaled s-, p- and d-shells on each atom.}

\vspace{1cm}

\captionof{table}{Total decay half-widths of the core-ionized neon atom, CIS-``out'' (reconstituted) calculated with different methods.}

{\begin{tabular}{lcccccc}\toprule
 & \multicolumn{3}{c}{CS/cc-pCV5Z} & \multicolumn{3}{c}{CBF/cc-pCVTZ (5sp) + 2$\times$(spd)}\\
Method & $\theta_\mathrm{opt}$ & $\frac{\mathrm{d}E}{\mathrm{d}\theta}$ & 
$-\mathrm{Im}(E)$ & $\theta_\mathrm{opt}$ & $\frac{\mathrm{d}E}{\mathrm{d}\theta}$ & 
$-\mathrm{Im}(E)$ \\
& / $^\circ$ & / Hartree/$^\circ$ & / meV & / $^\circ$ & / Hartree/$^\circ$ & / meV \\ \midrule
EOMIP-CCSD & 10 & $3.3\cdot10^{-5}$ & 102 & 17 & $3.8\cdot10^{-5}$ & 115 \\
$\Delta$CCSD & 12 & $3.8\cdot 10^{-6}$ & 94 & 22 & $5.4\cdot10^{-5}$ & 129 \\
out-ACP-CIS & 9 & $2.1\cdot 10^{-4}$ & 125 & 20 & $1.1\cdot10^{-4}$ & 107 \\
CIS (2p$^{-1}$) & 9 & $5.0\cdot10^{-5}$ & 29 & 20 & $2.5\cdot10^{-5}$ & 24 \\
CIS (2s$^{-1}$) & 9 & $6.2\cdot10^{-5}$ & 28 & 20 & $3.0\cdot10^{-5}$ & 42 \\ \midrule
Fano-EOM-CCSD~\textsuperscript{a} & -- & -- & 109 & -- & -- & 109 \\
Experiment~\textsuperscript{b} & -- & -- & 129 & -- & -- & 129 \\ \bottomrule
\end{tabular}}\\[1ex]
\begin{raggedright}
\textsuperscript{a} From Ref. 26.\\
\textsuperscript{b} From Ref. 56.\\[1ex]
\end{raggedright}

\vspace{1cm}

\newpage
    
\captionof{table}{Total decay half-widths of the core-ionized ammonia and 
methane molecules, calculated with different methods and the cc-pCVTZ (5sp) 
basis set with 2 complex scaled s-, p- and d-shells on each atom.}
{\begin{tabular}{lccc} \toprule
\multicolumn{4}{c}{Ammonia} \\ \midrule
Method & $\theta_\mathrm{opt}$/$^\circ$ & $\frac{\mathrm{d}E}{\mathrm{d}\theta}$ 
/ Hartree/$^\circ$ & $-\mathrm{Im}(E)$ / meV \\ \midrule
EOMIP-CCSD & 21 & $5.3 \cdot 10^{-6}$ & 56 \\
$\Delta$CCSD & 15 & $8.6 \cdot 10^{-6}$ & 55 \\
out-ACP-CIS & 23 & $4.7\cdot 10^{-5}$ & 53 \\ \midrule
\multicolumn{4}{c}{Methane} \\ \midrule
EOMIP-CCSD & 19 & $1.5 \cdot 10^{-5}$ & 37 \\
$\Delta$CCSD & 13 & $2.3 \cdot 10^{-5}$ & 38 \\
out-ACP-CIS & 15 & $3.2 \cdot 10^{-5}$ & 37 \\ \midrule
Fano-EOM-CCSD~\textsuperscript{a} & -- & -- & 33 \\
Experiment\textsuperscript{b} & {--} & {--} & 47 \\ \bottomrule
\end{tabular}}\\[1ex]
\begin{raggedright}
\textsuperscript{a} From Ref. 26.\\
\textsuperscript{b} From Ref. 58.\\[1ex]
\end{raggedright}

\vspace{1cm}
    
\includegraphics[width=\linewidth]{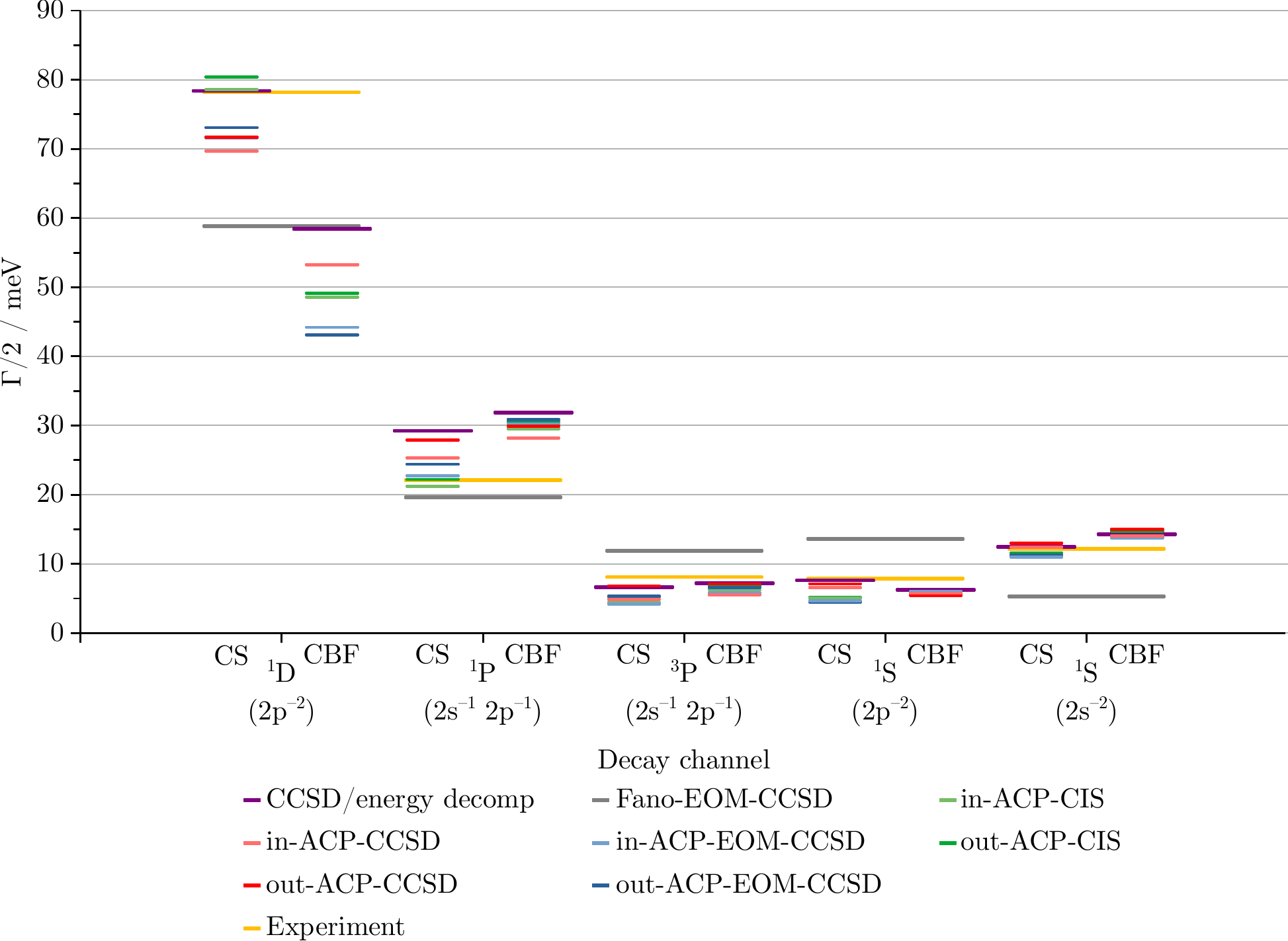}
\captionof{figure}{Partial widths of core-ionized neon, calculated with 
different methods. Experimental values from Ref. 56. Fano-EOM-CCSD 
values from Ref. 26.}

\newpage
 
\captionof{table}{Partial half-widths of core-ionized neon, calculated 
with different methods.}
\begin{tabular}{l|rrrrrrr|rr}\toprule
Decay & \multicolumn{2}{c}{CCSD} & \multicolumn{2}{c}{EOMIP-CCSD} 
& \multicolumn{2}{c}{CIS} & {CCSD} & {EOM-CCSD} & 
expt~\textsuperscript{b} \\
channel & \multicolumn{2}{c}{ACP} & \multicolumn{2}{c}{ACP} & \multicolumn{2}{c}{ACP} & 
energy & Fano~\textsuperscript{a} \\
 & ``in'' & ``out'' & ``in'' & ``out'' & ``in'' & ``out'' & decomp. & \\ \midrule
 & \multicolumn{7}{c|}{CS/cc-pCV5Z} & \\ \midrule
$^1$D (2p$^{-2}$) & 69.6 & 71.7 & 71.6 & 73.1 & 78.6 & 80.4 & 78.4 & 58.8 & 78.2(21) \\
$^1$P (2s$^{-1}$2p$^{-1}$) & 25.3 & 27.9 & 22.8 & 24.4 & 21.2 & 22.2 & 29.2 & 19.6 & 22.1(7) \\
$^3$P (2s$^{-1}$2p$^{-1}$) & 4.8 & 6.8 & 4.2 & 5.3 & 4.4 & 5.2 & 6.6 & 11.9 & 8.1(3) \\
$^1$S (2p$^{-2}$) & 6.6 & 7.1 & 4.7 & 4.5 & 5.1 & 5.1 & 7.6 & 13.6 & 7.9(3) \\
$^1$S (2s$^{-2}$) & 12.4 & 13.0 & 11.0 & 11.2 & 11.3 & 11.6 & 12.5 & 5.3 & 12.2(4) \\ \midrule
Sum & 118.9 & 126.5 & 114.2 & 118.5 & 120.6 & 124.5 & 134.3 & 109.2 & 128.5(30) \\\midrule
 & \multicolumn{7}{c|}{CBF/cc-pCVTZ (5sp) + 2$\times$(spd)} & \\ \midrule
$^1$D (2p$^{-2}$) & 56.3 & 56.7 & 44.2 & 43.1 & 48.6 & 49.1 & 58.4 & \\
$^1$P (2s$^{-1}$2p$^{-1}$) & 27.4 & 29.4 & 30.2 & 30.9 & 29.5 & 30.5 & 31.9 & \\
$^3$P (2s$^{-1}$2p$^{-1}$) & 5.3 & 7.2 & 5.8 & 6.6 & 6.2 & 6.9 & 7.2 & \\
$^1$S (2p$^{-2}$) & 5.4 & 5.1 & 6.0 & 5.7 & 5.8 & 5.8 & 6.2 &  \\
$^1$S (2s$^{-2}$) & 13.3 & 13.8 & 13.8 & 15.0 & 14.0 & 14.7 & 14.3 & \\ \midrule
Sum & 107.7 & 112.2 & 100.0 & 101.2 & 104.1 & 107.0 & 118.0 & \\ \midrule
 & \multicolumn{7}{c|}{CBF/cc-pCVTZ (5sp) + 4$\times$(spd)} & \\ \midrule
$^1$D (2p$^{-2}$) & -- & 69.4 & -- & 63.7 & -- & 68.0 & 75.5 \\
$^1$P (2s$^{-1}$2p$^{-1}$) & -- & 28.1 & -- & 25.8 & -- & 23.0 & 29.6 & \\
$^3$P (2s$^{-1}$2p$^{-1}$) & -- & 6.9 & -- & 5.6 & -- & 5.4 & 6.7 & \\
$^1$S (2p$^{-2}$) & -- & 4.9 & -- & 4.5 & -- & 3.1 & 5.7 & \\
$^1$S (2s$^{-2}$) & -- & 13.8 & -- & 12.3 & -- & 13.3 & 13.5 & \\ \midrule
Sum & -- & 123.1 & -- & 111.9 & -- & 112.8 & 131.0 & \\
\bottomrule
\end{tabular}\\[1ex]
\begin{raggedright}
\textsuperscript{a} From Ref. 26. \\
\textsuperscript{b} From Ref. 56. \\[1ex]
\end{raggedright}

\vspace{1cm}

\captionof{table}{Partial half-widths of core-ionized ammonia, calculated with 
different methods and the cc-pCVTZ (5sp) basis set with 2 complex scaled s-, p- 
and d-shells on each atom.} %NH3
\begin{tabular}{lrrrrrrr}\toprule
Decay & \multicolumn{2}{c}{CCSD} & \multicolumn{2}{c}{EOMIP-CCSD} & 
\multicolumn{2}{c}{CIS} & {CCSD} \\
channel & \multicolumn{2}{c}{ACP} & \multicolumn{2}{c}{ACP} & \multicolumn{2}{c}{ACP} & 
energy \\
 & ``in'' & ``out'' & ``in'' & ``out'' & ``in'' & ``out'' & decomp. \\ \midrule
$^1$A$_2$ (2a$_1^{-2}$) & 7.2 & 7.8 & 7.0 & 7.2 & 7.4 & 7.6 & 7.3 \\
$^1$E (2a$_1^{-1}$1e$^{-1}$) & 5.9 & 7.2 & 7.3 & 7.7 & 7.5 & 7.7 & 7.9 \\
$^3$E (2a$_1^{-1}$1e$^{-1}$) & 1.3 & 2.0 & 1.5 & 1.7 & 2.0 & 2.1 & 2.0 \\
$^1$A$_1$ (1e$^{-2}$) & 2.6 & 2.6 & 2.5 & 2.5 & 2.0 & 1.9 & 3.6 \\
$^1$E (1e$^{-2}$) & 9.8 & 10.2 & 8.2 & 8.3 & 9.4 & 9.5 & 11.7 \\
$^3$A$_2$ (1e$^{-2}$) & 0.0 & 0.0 & 0.1 & 0.0 & 0.0 & 0.0 & 0.0 \\
$^1$A$_1$ (3a$_1^{-2}$) & 5.6 & 5.8 & 4.7 & 4.8 & 5.7 & 5.8 & 6.7 \\
$^1$A$_1$ (2a$_1^{-1}$3a$_1^{-1}$) & 4.2 & 5.2 & 4.9 & 5.3 & 5.4 & 5.6 & 5.3 \\
$^3$A$_1$ (2a$_1^{-1}$3a$_1^{-1}$) & 0.7 & 1.1 & 0.8 & 1.0 & 1.0 & 1.1 & 1.1 \\
$^1$E (3a$_1^{-1}$1e$^{-1}$) & 11.6 & 12.2 & 10.1 & 10.3 & 11.3 & 11.5 & 14.2 \\
$^3$E (3a$_1^{-1}$1e$^{-1}$) & 0.1 & 0.1 & 0.1 & 0.1 & 0.2 & 0.2 & 0.1 \\ \midrule
Sum & 49.0 & 54.2 & 47.1 & 48.9 & 51.9 & 53.2 & 59.9 \\ \bottomrule
\end{tabular}\\[1ex]

\newpage

\includegraphics[width=\linewidth]{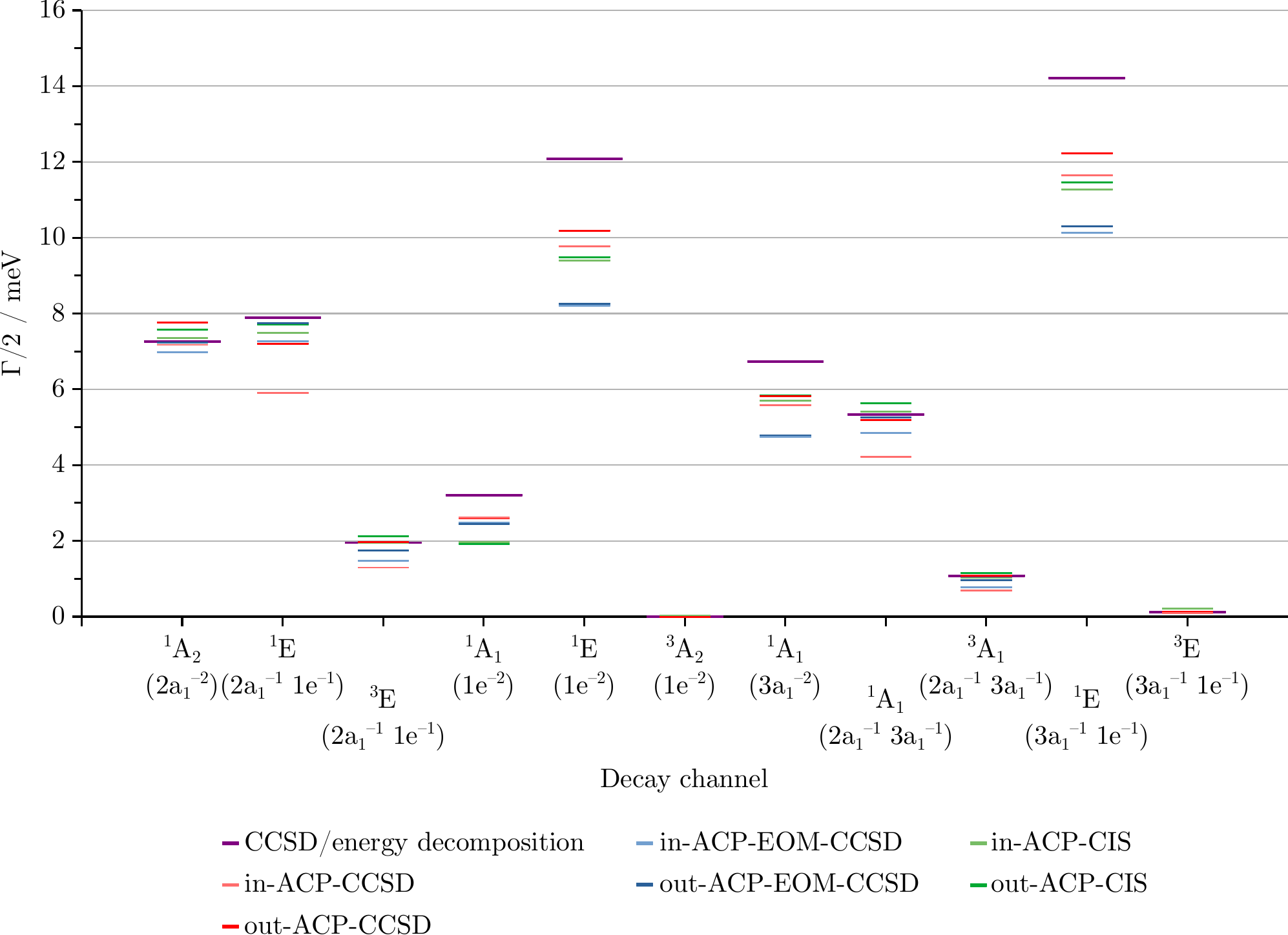}
\captionof{figure}{Partial widths of core-ionized ammonia, calculated with 
different methods and the cc-pCVTZ (5sp) basis set with 2 complex scaled 
s-, p- and d-shells on each atom.}

\vspace{1cm}

\captionof{table}{Partial half-widths of core-ionized methane, calculated with 
different methods and the cc-pCVTZ (5sp) basis set with 2 complex scaled 
s-, p- and d-shells on each atom.}%CH4
\begin{tabular}{lrrrrrrr|r}\toprule
Decay & \multicolumn{2}{c}{CCSD} & \multicolumn{2}{c}{EOMIP-CCSD} & 
\multicolumn{2}{c}{CIS} & CCSD &{EOM-CCSD} \\
channel & \multicolumn{2}{c}{ACP} & \multicolumn{2}{c}{ACP} & 
\multicolumn{2}{c}{ACP} & energy & Fano~\textsuperscript{a} \\
 & ``in'' & ``out'' & ``in'' & ``out'' & ``in'' & ``out'' & decomp. & \\ \midrule
$^1$A$_1$ (2a$_1^{-2}$) & 5.9 & 5.9 & 5.9 & 5.8 & 5.8 & 5.9 & 5.7 & 6.6 \\
$^1$T$_2$ (2a$_1^{-1}$1t$_2^{-1}$) & 6.7 & 8.6 & 8.3 & 8.7 & 6.7 & 7.1 & 9.5 & 7.1 \\
$^3$T$_2$ (2a$_1^{-1}$1t$_2^{-1}$) & 1.8 & 2.9 & 2.0 & 2.3 & 2.1 & 2.4 & 2.9 & 3.4 \\
$^1$T$_2$ (1t$_2^{-2}$) & 14.6 & 16.3 & 11.6 & 12.0 & 13.6 & 14.0 & 19.3 & 8.0 \\
$^1$A$_1$ (1t$_2^{-2}$) & 1.1 & 0.9 & 1.3 & 1.1 & 1.3 & 1.1 & 1.3 & 3.0 \\
$^1$E (1t$_2^{-2}$) & 7.9 & 8.3 & 6.5 & 6.5 & 6.3 & 6.3 & 9.9 & 5.4 \\
$^3$T$_1$ (1t$_2^{-2}$) & 0.0 & 0.0 & 0.0 & 0.0 & 0.1 & 0.1 & 0.0 & 0.0 \\ \midrule
Sum & 38.0 & 43.0 & 35.5 & 36.4 & 35.9 & 37.0 & 48.7 & 33.4 \\
\bottomrule
\end{tabular}\\[1ex]
\begin{raggedright}
\textsuperscript{b} From. Ref. 26. \\[1ex]
\end{raggedright}

\newpage

\includegraphics[width=\linewidth]{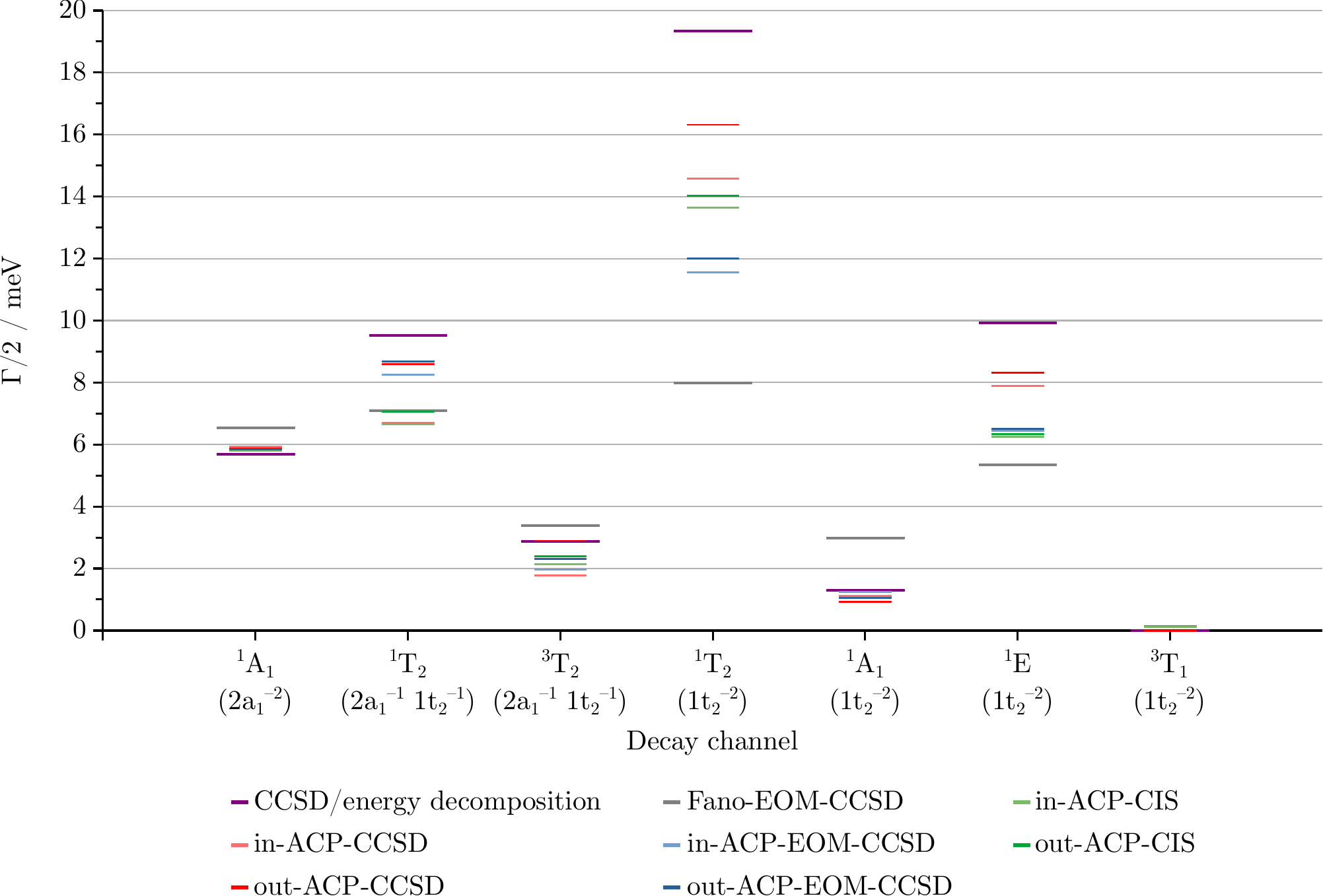}
\captionof{figure}{Partial widths of core-ionized methane, calculated with different methods and the cc-pCVTZ (5sp) basis set with 2 complex scaled s-, p- and d-shells on each atom. Fano-EOM-CCSD values from Ref. 26.}

\vspace{1cm}

\includegraphics[width=\linewidth]{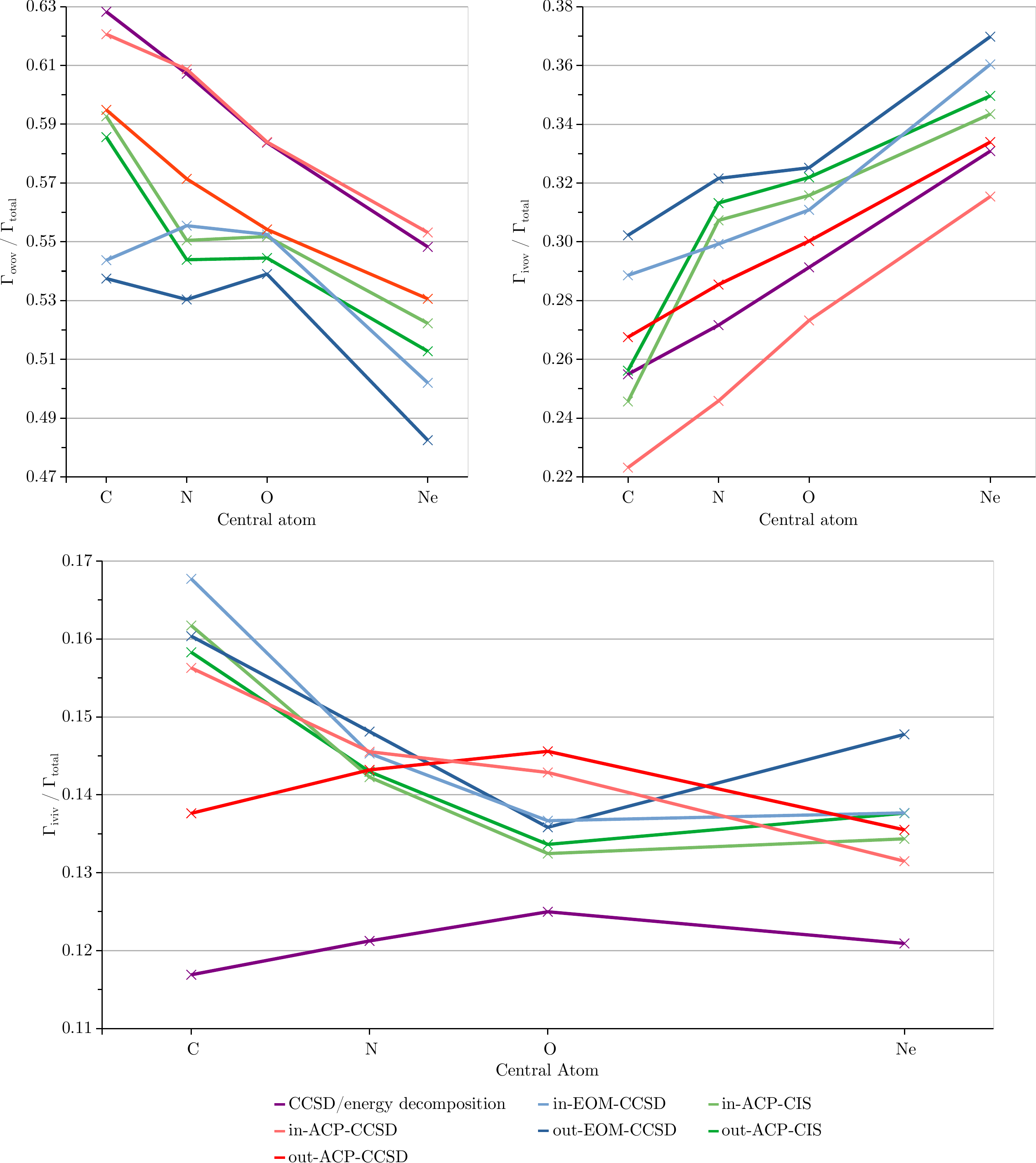}
\captionof{figure}{Relative contributions of decay channels involving two outer-valence spin-orbitals 
(upper left), one outer-valence and one inner-valence spin-orbital (upper right), 
and two inner-valence spin-orbitals (bottom) to the total Auger decay width of CH$_4$, 
NH$_3$, H$_2$O, and Ne.}